\documentclass[superscriptaddress, aps, prd, twocolumn, amsmath, amssymb, showpacs]{revtex4-2}

\usepackage{graphicx}  
\usepackage{mathptmx}
\usepackage{epstopdf}
\usepackage{dcolumn}   
\usepackage{bm}        
\usepackage{amssymb}   
\usepackage{amsmath}   
\usepackage{amsthm}
\usepackage{chemarrow}
\usepackage{color}
\usepackage[dvipsnames]{xcolor}
\usepackage{mathrsfs}
\usepackage{float}
\usepackage{physics}
\usepackage{slashed}
\usepackage{bbold}
\usepackage{bbm}
\usepackage{latexsym}

\usepackage[colorlinks=true, linkcolor=blue, citecolor=purple]{hyperref}

\usepackage{orcidlink}

\let\oldfootnote\footnote
\renewcommand{\footnote}[1]{%
    \begingroup%
    \linespread{1}
    \oldfootnote{#1}%
    \endgroup%
}
\newcommand{\beq}{\begin{equation}}
\newcommand{\eeq}{\end{equation}}
\newcommand{\beqa}{\begin{eqnarray}}  
\newcommand{\eeqa}{\end{eqnarray}}  
\newcommand{\bit}{\begin{itemize}}  
\newcommand{\eit}{\end{itemize}}  

\newcommand{\eg}{{\it e.g.}}

\newcommand{\Gsub}{G_{\mu\nu}}
\newcommand{\Gtildesup}{\widetilde{G}^{\mu\nu}}

\newcommand{\fa}{f}
\newcommand{\fpi}{f_\pi}
\newcommand{\qbar}{\overline{q}}
\newcommand{\kLo}{\mathbf{k}_{L,0}}
\newcommand{\kRo}{\mathbf{k}_{R,0}}
\newcommand{\kL}{\mathbf{k}_L}
\newcommand{\kR}{\mathbf{k}_R}

\newcommand{\QLo}{\boldsymbol{\theta}_{L,0}}
\newcommand{\QRo}{\boldsymbol{\theta}_{R,0}}

\newcommand{\kappaqo}{\boldsymbol{\kappa}_{q,0}}

\newcommand{\deltaqo}{\boldsymbol{\delta}_{q,0}}

\newcommand{\deltacud}{\Delta c_{ud}}
\newcommand{\dkappao}{\delta_{\kappa,0}}
\newcommand{\cA}{\mathcal{A}}

\allowdisplaybreaks

\begin{document}

\title{Wess-Zumino-Witten Interactions of Axions}

\author{Yang Bai \orcidlink{0000-0002-2957-7319}}
\affiliation{
Department of Physics, University of Wisconsin-Madison, Madison, WI 53706, USA}

\author{Ting-Kuo Chen \orcidlink{0000-0002-5267-6729}}
\affiliation{
Department of Physics, University of Wisconsin-Madison, Madison, WI 53706, USA}

\author{Jia Liu \orcidlink{0000-0001-7386-0253}}
\affiliation{School of Physics and State Key Laboratory of Nuclear Physics and Technology, Peking University, Beijing 100871, China}
\affiliation{Center for High Energy Physics, Peking University, Beijing 100871, China}

\author{Xiaolin Ma \orcidlink{0009-0007-1994-9493}}
\affiliation{School of Physics and State Key Laboratory of Nuclear Physics and Technology, Peking University, Beijing 100871, China}

\begin{abstract}
We present a consistent derivation of the complete Wess-Zumino-Witten interactions of axions, including the counter-term necessary to guarantee the gauge invariance of the Standard Model. By treating the derivative of the axion field as a background gauge field and incorporating auxiliary chiral rotation phases, we ensure consistency in the axion-interaction Lagrangian. This approach allows us to derive basis-independent physical interactions of axions with gauge bosons and vector mesons. As an example, we explore the interaction of $a$-$\omega$-$\gamma$ to illustrate the potential for searching for axion-like particles at colliders.  
\end{abstract}

\maketitle

\noindent \textbf{Introduction}. Axions and axion-like particles (ALPs) are pseudo-Nambu-Goldstone bosons resulting from the spontaneous breaking of a $U(1)$ global symmetry. They are ubiquitously predicted not only in particle physics, where they serve as a solution to the strong-CP problem~\cite{Peccei:1977hh, Peccei:1977ur, Weinberg:1977ma, Wilczek:1977pj}, but also in string theory~\cite{Svrcek:2006yi}. Axions could be non-thermally produced in the early universe and potentially account for all or part of the dark matter~\cite{ParticleDataGroup:2022pth}.  

The search for axion particles relies on their couplings to the Standard Model (SM) fermions and gauge bosons. The derivation of the axion coupling to two photons requires evolving the theory from a scale above the QCD scale to a scale below it. This process requires special care to track both the global-gauge-gauge mixed anomaly and the mass and kinetic mixings between the axion and meson fields. For instance, Refs.~\cite{Bauer:2020jbp, Bauer:2021wjo} introduced auxiliary chiral rotation phases and demonstrated \textit{consistent} physical results for both $a \rightarrow \gamma\gamma$ and $K \to \pi a$ processes that are independent of the auxiliary chiral phases.

Beyond the well-studied axion-photon interaction, axion interactions with other pseudo-scalar or (axial-)vector mesons are universally predicted in many axion models and may provide an important avenue to search for axions. These interactions may involve global-global-global or global-global-gauge mixing anomalies, with the 't Hooft anomaly matching~\cite{tHooft:1979rat} playing an essential role. For instance, the interaction of $a$-$\omega$-$\gamma$ related to the $U(1)_a$-$U(1)_{\rm baryon}$-$U(1)_{\rm EM}$ anomaly requires consistent matching from above to below the QCD scale (see Ref.~\cite{Metlitski:2005pr} for an earlier study). However, it is unclear how to obtain the complete and consistent axion interactions with various fundamental gauge bosons like the photon and the neutral $Z$ boson, as well as composite pseudo-scalar and (axial-)vector mesons. 

It is well known that the interactions of pseudo-scalar mesons with fundamental gauge bosons can be derived using the Wess-Zumino-Witten (WZW) term~\cite{Wess:1971yu, Witten:1983tw}. The WZW term represents a class of topological operators, which can be conceptualized as the boundary term of a $D=5$ manifold to reflect the anomaly structure of QCD. When coupled to the background gauge fields, it can reproduce the anomalies of the underlying quark theories as a consequence of the 't Hooft anomaly matching. Recently, Refs.~\cite{Harvey:2007rd, Harvey:2007ca} have extended the WZW term to incorporate the electroweak $SU(2)_L \times U(1)_Y$ gauge fields and (axial-)vector mesons as background fields coupling to various global currents. To maintain the $SU(2)_L \times U(1)_Y$ gauge invariance, a counter-term is required to be added to the WZW term. The summation of the WZW term and the counter-term leads to new physical interactions, termed pseudo-Chern-Simons terms in Refs.~\cite{Harvey:2007rd, Harvey:2007ca}.

\begin{figure}[t!]
\centering
\includegraphics[width= 1 \linewidth]{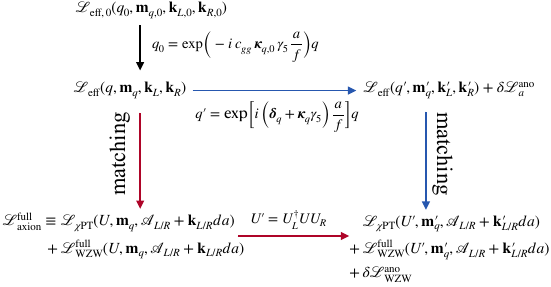}
\caption{Relations of the effective Lagrangians at the quark level (upper) and the matched chiral Lagrangians at the meson level (lower). The first quark chiral rotation from $q_0$ to $q$ is to eliminate the $a G\tilde{G}$ term from $\mathcal{L}_{\rm eff}$. The second chiral rotation from $q$ to $q'$ by introducing the auxiliary phases $\boldsymbol{\delta}_q$ and $\boldsymbol{\kappa}_q$ is for the consistent check of our derivation of the full axion interactions in the lower left corner. The two routes (labelled by the blue and red arrows) yield consistent results because the axion anomalous interactions at the quark level, $\delta \mathcal{L}^{\rm ano}_a$, matches the corresponding WZW anomalous term, $\delta \mathcal{L}^{\rm ano}_{WZW}$, from the chiral rotation at the meson level.
}
\label{fig:two-routes} 
\end{figure}

In this work, we present a \textit{consistent} derivation of the complete WZW interactions of axions, which can describe all axion interactions with pseudo-scalar mesons, fundamental gauge bosons, and composite vector mesons. We treat the derivative of the axion field as a background gauge field to incorporate the axion field into the WZW term plus counter-term. Similar to the treatment in Ref.~\cite{Bauer:2021wjo}, we introduce some auxiliary chiral rotation phases to perform a consistent check on the complete axion-interaction Lagrangian, which includes both the normal chiral Lagrangian and the full WZW interaction Lagrangian (see the lower left corner of Fig.~\ref{fig:two-routes}). Equipped with the consistent WZW interactions of axions, we then choose the $a$-$\omega$-$\gamma$ interaction as an example to illustrate the phenomenological consequences of the full WZW interactions of axions.
\\

\noindent \textbf{Effective Lagrangian}.
We begin by defining the SM gauge bosons (denoted as \(\mathbb{A}\)) and background fields (denoted as \(\mathbb{B}\)) in their one-forms, \(\mathbb{A}(\mathbb{B})_{L,R}\equiv \mathbb{A}(\mathbb{B})_{L,R}^\mu dx_\mu\), to simplify the notation. The SM electroweak gauge boson fields can be expressed as
\begin{equation}
    \mathbb{A}_L = \frac{e}{s_{\rm w}}W^i\frac{\boldsymbol{\tau}^i}{2} + \frac{e}{c_{\rm w}}W^0\mathbf{Y}_Q ~, 
   \quad  \mathbb{A}_R = \frac{e}{c_{\rm w}}W^0\mathbf{Y}_q ~,
\label{eq:gauge-forms}
\end{equation}
where $W^i$ and $W^0$ stand for the $SU(2)_L$ and $U(1)_Y$ gauge one-forms, respectively. Here, $s_{\rm w}\equiv \sin{\theta_{\rm w}}$ and $c_{\rm w}\equiv \cos{\theta_{\rm w}}$ with $\theta_{\rm w}$ as the weak mixing angle, $\mathbf{Y}_Q={\rm diag}(1/6,1/6)$, and $\mathbf{Y}_q={\rm diag}(2/3,-1/3)$. The (axial-)vector meson background fields are given by~\cite{Harvey:2007ca} 
\begin{equation}
\begin{aligned}
    \mathbb{B}_V &\equiv \mathbb{B}_L + \mathbb{B}_R =   g\begin{pmatrix}
        \rho_0 & \sqrt{2}\,\rho^+ \\
        \sqrt{2}\,\rho^- & -\rho_0
    \end{pmatrix} + g'\begin{pmatrix}
        \omega & \\
        & \omega
    \end{pmatrix}  ~, \\
    \mathbb{B}_A &\equiv \mathbb{B}_L - \mathbb{B}_R= g\begin{pmatrix}
        a_1 & \sqrt{2}\,a^+ \\
        \sqrt{2}\,a^- & -a_1
    \end{pmatrix} + g'\begin{pmatrix}
        f_1 & \\
        & f_1
    \end{pmatrix} ~.
\end{aligned}
\label{eq:vec_meson_fields11}
\end{equation}
Above the QCD scale, these (axial-)vector meson fields should be thought of as fictitious background gauge fields used to keep track of the global-global-gauge and global-global-global anomalies.
As we will show later, the axion field should be incorporated into the background fields in order for the physical observables to be consistent, 
\begin{equation}
    \mathbb{B}_{L/R} \to \mathbb{B}_{L/R} + \mathbf{k}_{L/R,0}\,da/\fa ~.
\end{equation}
Consequently, we can write down the general axion effective Lagrangian above the QCD and below the electroweak scale 
\begin{align}
    \mathcal{L}_{\rm eff,0} &= \mathcal{L}_{\rm SM} +  \bar{q}_0(i\slashed{D}-\mathbf{m}_{q,0})q_0 + \frac{1}{2}(\partial_\mu a)(\partial^\mu a) - \frac{m_{a,0}^2}{2}a^2   \nonumber \\
    + & \, c_{gg}\frac{\alpha_s}{4\pi}\frac{a}{\fa}\Gsub\Gtildesup + \frac{a}{\fa}\sum_{\cA_{1,2}} c_{\cA_{1} \cA_{2}}^0 F_{\cA_{1} \,\mu\nu}\,\widetilde{F}_{\cA_{2}}^{\mu\nu} +\mathcal{L}_c ~,
    \label{eq:ALP-eff-quark-initial}
\end{align}
where $\cA=\mathbb{A},\mathbb{B}$. $F_\cA$ denotes the corresponding field strength, and $\widetilde{F}_{\mu\nu}=\frac{1}{2}\epsilon^{\mu\nu\rho\sigma}F_{\rho\sigma}$.
The boldfaced symbols $\mathbf{m}, \mathbf{k}$ are matrices in the flavor space.
For simplicity, we focus on two-flavor quarks: up-quark and down-quark with $q_0 \equiv (u_0, ~d_0)^T$ (see Ref.~\cite{Bauer:2021mvw} for the case with more flavors). In $\mathcal{L}_{\rm eff,0}$, the quark mass matrix is real, diagonal, and does not couple to axions, defined as \(\mathbf{m}_{q,0} = {\rm diag}(m_u, ~m_d)\).

Note that the covariant derivative also contains the axion field, which thus encodes the derivative coupling of the axion to the chiral quark currents~\cite{Gasser:1984gg}, and is given by
\begin{equation}
    D^\mu = \partial^\mu - i\sum_{\cA}g_{\cA}\left(\cA_{L}^\mu P_L + \cA_{R}^\mu P_R\right) ~,
\end{equation}
where $g_\cA$ denotes the gauge coupling and $P_{L, R} = (1 \mp \gamma_5)/2$. In the presence of background gauge fields, the counter-term $\Gamma_c = \int d^4 x \mathcal{L}_c (\mathbb{A}_{L/R},\mathbb{B}_{L/R})$ is needed to preserve the fundamental gauge invariance~\footnote{For vector-like fundamental gauge bosons, the counter-term is called the Bardeen counter-term. For chiral fundamental gauge bosons, the counter-terms are worked out in Ref.~\cite{Harvey:2007ca}.}
\begin{align}
	& \Gamma_c = -2\mathcal{C}\int \Tr\Big[ (\mathbb{A}_Ld\mathbb{A}_L+d\mathbb{A}_L\mathbb{A}_L)\mathbb{B}_L + \frac{1}{2}\mathbb{A}_L(\mathbb{B}_Ld\mathbb{B}_L+d\mathbb{B}_L\mathbb{B}_L) \nonumber \\
	&   - \frac{3}{2}\,i\,\mathbb{A}_L^3\mathbb{B}_L  - \frac{3}{4}\,i\,\mathbb{A}_L\mathbb{B}_L\mathbb{A}_L\mathbb{B}_L - \frac{i}{2}\mathbb{A}_L\mathbb{B}_L^3 \Big] - (L\leftrightarrow R) ~,
 \label{eq:WZW:counterterm1}
\end{align}
with $\mathcal{C} = N_c/(48\pi^2)$.

To match the standard chiral Lagrangian, we perform an axion-dependent phase rotation of the quark field,
\begin{equation}\label{eq:chiral:0}
    q_0 = \exp\left(-i\,c_{gg}\,\kappaqo \,\gamma_5\,\frac{a}{f} \right)q ,\quad \mbox{with}\;\Tr\left(\kappaqo\right) = 1 ~,
\end{equation}
which effectively shifts $\mathbf{m}_{q,0} \to \mathbf{m}_{q}(a)$ and $\mathbf{k}_{L/R,0} \to \mathbf{k}_{L/R}(a)$,
to arrive at a Lagrangian without the $a G \widetilde{G}$ term,
\begin{align}
    \mathcal{L}_{\rm eff} &= \mathcal{L}_{\rm SM} + \qbar i\slashed{D} q - \left[\qbar_L \,\mathbf{m}_q(a)\,q_R + h.c.\right] + \frac{1}{2}(\partial_\mu a)(\partial^\mu a)  \nonumber \\
    & - \frac{m_{a,0}^2}{2}a^2 + \frac{a}{\fa}\sum_{\cA_{1,2}} c_{\cA_1\cA_2} F_{\cA_1\,\mu\nu}\,\widetilde{F}_{\cA_2}^{\mu\nu}+ \mathcal{L}_c  
~.
\label{eq:ALP-eff-quark-noGG}
\end{align}
The couplings in this new basis are related to those in Eq.~\eqref{eq:ALP-eff-quark-initial} and are presented in Eq.~\eqref{eq:Leff:couplings}. 
Note that the physical observables will only depend on $\Tr\left(\kappaqo\right)$ and be independent of the individual entries of $\kappaqo$. This basis independence has been emphasized and explicitly checked in Ref.~\cite{Bauer:2021wjo}. This strategy will be adopted in our analysis to identify the basis-independent physical observables for the axion WZW interactions. 

In this study, we are also interested in deriving the axion couplings to chiral gauge bosons and composite (axial-)vector mesons in the SM. To check the basis-independent physical observables, we introduce 3 independent auxiliary phases: $\boldsymbol{\delta}_q = \mbox{diag}(\delta_u, \delta_d)$ and $\boldsymbol{\kappa}_q = \mbox{diag}(\kappa_u, \kappa_d)$ subject to the constraint $\Tr(\boldsymbol{\kappa}_q)=0$ to avoid regenerating the $a\,G\widetilde{G}$ term, as in Ref.~\cite{Bauer:2021wjo}.
The quark fields in the two bases are related by
\beqa
\label{eq:two-basis-rotation}
q^\prime =  \exp \Bigl[  i\left(\boldsymbol{\delta}_q +\boldsymbol{\kappa}_q  \gamma_5 \right) a/f \Bigr]  q ~.
\eeqa
Defining $\boldsymbol{\theta}_{L/R} \equiv \boldsymbol{\delta}_q \mp  \boldsymbol{\kappa}_q$ and $U_{L/R} \equiv  \exp \bigl[  -i \boldsymbol{\theta}_{L/R}   a/f \bigr]$, the quark mass and quark bilinear couplings transform as
\begin{align}
\mathbf{m}'_q = U_L^\dagger \mathbf{m}_q U_R , \quad 
\mathbf{k}'_{L/R}= U_{L/R}^\dagger(\mathbf{k}_{L/R} + \boldsymbol{\theta}_{L/R} )U_{L/R} ~.
\label{eq:m-and-kLR}
\end{align}

From the $q$ basis to the $q'$ basis, the effective Lagrangian is transformed to be 
\beqa
\mathcal{L}_{\rm eff}(q, \mathbf{m}_q, \mathbf{k}_L,\mathbf{k}_R) \rightarrow \mathcal{L}_{\rm eff}(q', \mathbf{m}'_q, \mathbf{k}'_L,\mathbf{k}'_R) + \delta \mathcal{L}^{\rm ano}_{a} ~,
\label{eq:delta-L-ano-a}
\eeqa
with $\delta \mathcal{L}^{\rm ano}_{a}$ tracking the additional axion anomalous interactions including the familiar axion coupling to two photons as well as its other couplings to the background gauge bosons. This term is equivalent to the variation of the full WZW term below the QCD scale [see Eq.~\eqref{eq:delta:WZW}], in accordance with the requirement of 't Hooft anomaly matching~\cite{Harvey:2007ca}.
In the top line of Fig.~\ref{fig:two-routes}, we show the blue arrow to relate the two quark-level $\mathcal{L}_{\rm eff}$'s. In the vertical direction, we will match the quark-level Lagrangians to the chiral Lagrangians including the WZW interactions and demonstrate the agreement between the two routes through which we derive the Lagrangian at the lower-right corner starting from the Lagrangian at the upper-left corner. After that, we will use a few physical observables to demonstrate the independence of the auxiliary parameters $\boldsymbol{\delta}_{u,d}$ and $\boldsymbol{\kappa}_{u,d}$.
\\

\noindent \textbf{Chiral Lagrangian}.
Following the standard procedure to derive the chiral Lagrangian~\cite{Gasser:1984gg, Bauer:2021wjo} from Eq.~\eqref{eq:ALP-eff-quark-noGG}, one has 
\begin{align}
   & \mathcal{L}_{\chi{\rm PT}}  = \frac{\fpi^2}{8} \Tr \left[(D^\mu U)(D_\mu U)^\dagger\right] + \frac{\fpi^2}{4}B_0 \Tr [\mathbf{m}_q(a)\, U^\dagger + h.c.] \nonumber \\
  &\;+ \frac{1}{2}(\partial_\mu a)(\partial^\mu a) - \frac{m_{a,0}^2}{2}a^2 \,+ \, \frac{a}{\fa}\sum_{\cA_{1,2}} c_{\cA_1\cA_2} F_{\cA_1\,\mu\nu}\,\widetilde{F}_{\cA_2}^{\mu\nu} ~,
  \label{eq:chiral-L1}
\end{align}
where we leave the counter-term $\mathcal{L}_c$ aside from the chiral Lagrangian $\mathcal{L}_{\chi{\rm PT}}$ and will include it together with the WZW term later.
Below the QCD scale, one has dynamical $\omega$ and $\rho$ vector mesons that couple to the baryon-number and isospin currents, respectively. So, the global-global-gauge anomaly coefficient like $U(1)_a$-$U(1)_{\rm baryon (isospin)}$-$U(1)_{\rm EM}$ becomes the anomaly coefficient, $c_{(\omega/\rho)\gamma}$, of the axion coupling to $\omega/\rho$ vector meson and $\gamma$. The same is true for the couplings $c_{\omega\omega}$ and $c_{\rho\rho}$ that come from the global-global-global anomalies. For our later presentation convenience, we define the last term as
\beqa
\mathcal{L}_{\chi{\rm PT}}^{\rm ano} \equiv \frac{a}{\fa}\sum_{\cA_{1,2}} c_{\cA_1\cA_2} F_{\cA_1\,\mu\nu}\,\widetilde{F}_{\cA_2}^{\mu\nu} 
\label{eq:chi-PT-ano}
\eeqa
to include both UV-dependent as well as the chiral-rotation-generated axion anomalous couplings to vector bosons.

Here, $U=\exp[(\sqrt{2}i/\fpi)\pi^a\boldsymbol{\tau}^a]$ describes the Goldstone pion fields of the \(SU(2)_L\times SU(2)_R\) symmetry and \(\tau^a\) (with $\mbox{Tr}(\tau^a \tau^b) = 2 \delta^{ab}$) denotes the generators of the broken \(SU(2)_A\). The pion decay constant is \(\fpi\approx130\) MeV, and \(B_0\approx m_\pi^2/(m_u+m_d)\). The mass term \(\mathbf{m}_q(a)\) here contains axion-dependent phases. 
As in the quark-level case, the derivative coupling of the axion to the chiral quark currents can be incorporated into the definition of the pion matrix covariant derivative~\cite{Gasser:1984gg, Harada:2003jx, Bauer:2021wjo}
\begin{equation}
\label{eq:covariant-derive-with-axion}
\begin{aligned}
D^\mu U &= \partial^\mu U - i\sum_\cA g_\cA \left[\mathcal{A}_L^{\mu}U-U \mathcal{A}_R^{\mu}\right] ~,
\end{aligned}
\end{equation}
where $\mathcal{A}_{L/R}$ represents the left- and right-handed chiral gauge bosons (including both fundamental and emerged composite gauge bosons) and $g_{\cA}$ denotes the gauge couplings~\footnote{Using the aforementioned chiral Lagrangian, Refs.~\cite{Bauer:2020jbp, Bauer:2021wjo} have demonstrated consistent results for pseudoscalar meson decay to axion, such as \(K^- \to \pi^- a\) and \(\pi^- \to e^- \bar{\nu}_e a\), as well as the axion decay \(a \to \gamma \gamma\). These results are independent of auxiliary chiral phases. Later, we will show that the results are consistent because the WZW interactions do not contribute to these processes. On the other hand, the WZW interactions do contribute to other couplings like $a$-$\omega$-$\gamma$ and $a$-$\rho$-$\gamma$.
}. 
\\

\noindent \textbf{WZW interactions of axions}.
The WZW term is a crucial component of the chiral Lagrangian, closely tied to the anomaly structure of QCD. Incorporating (axial-)vector background fields into the WZW term offers a systematic approach to handle (axial-)vector meson interactions within chiral Lagrangian calculations~\cite{Harvey:2007rd, Harvey:2007ca}. In the literature, the global anomaly structure of the underlying axion-quark interactions is often overlooked. To obtain a complete anomaly description, we will systematically embed axions into the WZW interactions by treating its one-form ($da$) as a vector background field for the Peccei-Quinn symmetry. This approach will lead to new interactions among axions, (chiral/vector-like) fundamental gauge bosons, and background or composite (axial-)vector fields. Such interactions are essential for obtaining \textit{consistent} amplitudes for physical processes, a feat not achieved in previous literature on axions. 

To begin, we refer again to Eqs.~\eqref{eq:gauge-forms} and \eqref{eq:vec_meson_fields11} for the definitions of the vector one-forms.
Combining the fundamental gauge bosons and background/composite vector fields as \(\mathcal{A}_{L/R} \equiv \mathbb{A}_{L/R} + \mathbb{B}_{L/R}\), the full fundamental-gauge-invariant WZW interactions are given by~\cite{Harvey:2007ca}
\begin{align}
    \mathcal{L}_{\rm WZW}^{\rm full}(U, \mathcal{A}_{L/R}) = \mathcal{L}_{\rm WZW}(U,\mathcal{A}_L, \mathcal{A}_R) + \mathcal{L}_c(\mathbb{A}_{L/R}, \mathbb{B}_{L/R}) ~,
\end{align}
with the explicit formula of $\mathcal{L}_{\rm WZW}(U,\mathcal{A}_L, \mathcal{A}_R)$ given by Eq.~\eqref{eq:WZW} in the Appendix. 

Next, as in the quark-level case, we include axions into the WZW interactions by adding the axion one-form $d a$ into the background fields given the form of $D^\mu U$ in Eq.~\eqref{eq:covariant-derive-with-axion} 
\begin{align}
    \mathbb{B}_{L/R} &\to \mathbb{B}_{L/R} + \mathbf{k}_{L/R}(a)\,da/\fa ~, \nonumber \\ 
    \mathcal{L}_{\rm WZW}^{\rm full}(U, \mathcal{A}_{L/R}) &\to \mathcal{L}_{\rm WZW}^{\rm full}(U, \mathcal{A}_{L/R} + \mathbf{k}_{L/R}(a)\, da/\fa) ~.
    \label{eq:axion-WZW-full}
\end{align}
The summation of $\mathcal{L}_{\chi{\rm PT}}$ from Eq.~\eqref{eq:chiral-L1} and $\mathcal{L}_{\rm WZW}^{\rm full}(U, \mathcal{A}_{L/R} + \mathbf{k}_{L/R}(a)\, da/\fa)$ provides the complete interactions for the axion field with both pseudoscalar and vector fields (shown in the lower-left corner of Fig.~\ref{fig:two-routes}), or, 
\beq
\label{eq:chiral-Lag-full} 
\mathcal{L}_{\rm axion}^{\rm full} 
\equiv \left[\mathcal{L}_{\rm \chi PT}+ \mathcal{L}_{\rm WZW}^{\rm full}\right]\left(U,\mathbf{m}_q(a), \mathcal{A}_{L/R} + \mathbf{k}_{L/R}(a) da/\fa\right) .
\eeq

Coming back to \(\mathcal{L}_{\rm WZW}^{\rm full}\) or the axion WZW interactions, we have systematically worked out the relevant terms for the axion couplings. We categorize them into three-point and four-point vertex forms, denoted as \(XdYda\) and \(XYZda\), respectively. Here, \(X/Y/Z\) represent the SM gauge bosons or (axial-)vector meson fields. The complete expressions for the axion-relevant WZW interactions, \(\Gamma^{XdYda}_{\rm WZW}\) and \(\Gamma^{XYZda}_{\rm WZW}\), are provided in Appendix~\ref{sec:axion-WZW-terms}. Note that $\gamma d\gamma da$ from the WZW interactions vanishes due to its vector-like gauging and the presence of the counter-term~\cite{Harvey:2007ca}.
\\

\noindent \textbf{Matching between $\mathcal{L}_{\rm eff}$ and $\mathcal{L}_{\rm axion}^{\rm full} $}.
We can use the two different routes (blue and red arrows in Fig.~\ref{fig:two-routes}) to derive the same Lagrangian at the lower-right corner to demonstrate the consistency of our derivation of the full Lagrangian. Using the rotation in Eq.~\eqref{eq:two-basis-rotation} from \(q\) to \(q'\), one has the corresponding transformations for the meson matrix, fundamental gauge fields and background/composite meson fields
\beqa
 U =  U_L U' U_R^\dagger\,, \quad 
\delta \mathbb{A}_{L/R}=0\,, \quad 
\delta \mathbb{B}_{L/R}=\boldsymbol{\theta}_{L/R}da/\fa ~.
\eeqa
Note that \(\mathbf{k}'_{L/R} = \mathbf{k}_{L/R} + \boldsymbol{\theta}_{L/R} \) due to the fact that the relevant matrices are all diagonal. 
The anomalous part of the WZW interactions resulting from this auxiliary chiral rotation, \(\delta \mathcal{L}_{\rm WZW}^{\rm ano}\), can be expressed as
\beqa\label{eq:delta:WZW}
     \delta \mathcal{L}_{\rm WZW}^{\rm ano} & =& \mathcal{L}_{\rm WZW}^{\rm full}\left(U, \mathbf{m}_q(a), \mathcal{A}_{L/R} + \mathbf{k}_{L/R}(a) da/\fa\right)  \nonumber \\
   & & \hspace{-3.0mm} - \mathcal{L}_{\rm WZW}^{\rm full}\left(U', \mathbf{m}'_q(a), \mathcal{A}_{L/R} + \mathbf{k}'_{L/R}(a) da/\fa\right) ~. 
\eeqa
The explicit formula of $\delta \mathcal{L}_{\rm WZW}^{\rm ano}$ can be found in Eq.~\eqref{eq:anomalous} and is coincident to the term $\delta \mathcal{L}^{\rm ano}_{a}$ in Eq.~\eqref{eq:delta-L-ano-a}, since both originate from the same chiral phase rotation (either in terms of quark or meson fields). This completes our consistent proof that the Lagrangian in the lower-left corner of Fig.~\ref{fig:two-routes} or $\mathcal{L}_{\rm axion}^{\rm full}$ in Eq.~\eqref{eq:chiral-Lag-full} is basis-independent~\footnote{Note that the notations between Ref.~\cite{Harvey:2007ca} and Ref.~\cite{Kaymakcalan:1983qq} have an overall sign difference for the WZW interactions.}. 
\\

\noindent \textbf{Consistent physical interactions}.
One can use the full axion Lagrangian to derive various physical interactions and demonstrate their independence of the 3 auxiliary phases. In Appendix~\ref{app:physical-amplitude}, we have chosen a few examples as demonstrations, $a$-$\gamma$-$\gamma$, $a$-$\omega$-$\gamma$, $a$-$\rho$-$\gamma$, $a$-$Z$-$\gamma$ and $a$-$Z$-$\omega$. While the consistent derivation of $a$-$\gamma$-$\gamma$ has already been performed in Ref.~\cite{Bauer:2020jbp}, we have confirmed this as well in Appendix~\ref{app:a-gamma-gamma} based on the full axion Lagrangian in Eq.~\eqref{eq:chiral-Lag-full}. 

\begin{figure}[ht!]
\centering
\includegraphics[width= 0.99 \linewidth]{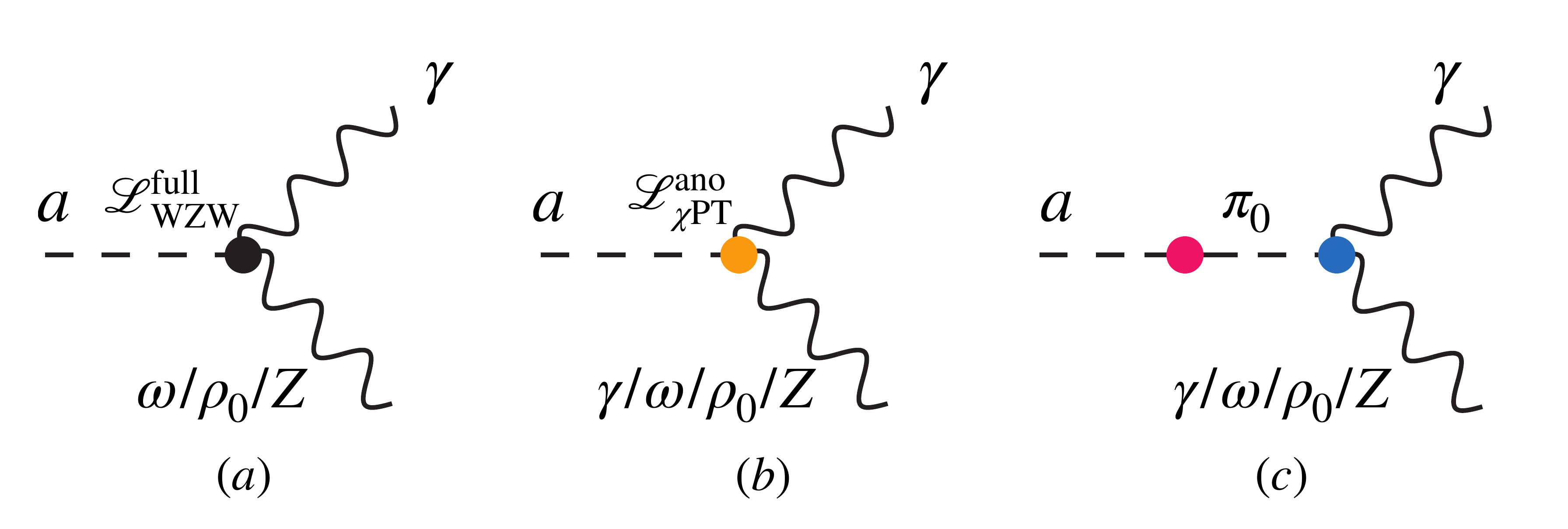}
\caption{The relevant Feynman diagrams depicting the effective coupling of \(a\)-\(\gamma/\omega/\rho_0/Z\)-\(\gamma\): (a) from the full WZW interactions in Eq.~\eqref{eq:axion-WZW-full}; (b) from the interaction $\mathcal{L}_{\rm \chi PT}^{\rm ano}$ in Eq.~\eqref{eq:chi-PT-ano}; (c) from the mass and kinetic mixings between the axion and the pion.} 
\label{fig:Feynman-diagrams} 
\end{figure}

For the axion couplings with one photon and one vector meson $\omega$ or $ \rho_0$, we have performed a basis-independent consistency check in Appendix~\ref{app:a-omega-gamma}. For the UV model with $c_{\cA_{1} \cA_{2}}^0 = 0$, the physical coefficient for the $a d\omega d\gamma$ interaction with all particles on-shell is given by
\begin{align}
  c_{\omega\gamma}^{\rm eff} 
&= eg'\left\{\frac{-c_{gg}}{8\pi^2f} +\frac{1}{16\pi^2f}(c_Q-2c_u+c_d) \right. \label{eq:omegagamma}
 \\
&\left.\hspace{-2mm} -\frac{3}{8\pi^2f}\left[\frac{m_a^2}{m_\pi^2-m_a^2}\left(\frac{c_{u}-c_{d}}{2}\right)+c_{gg}\frac{m_u-m_d}{m_u+m_d}\frac{m_\pi^2}{m_a^2-m_\pi^2}\right]\right\} ~, \nonumber 
\end{align}
with the corresponding formula for $c_{\rho\gamma}^{\rm eff}$ given in Eq.~\eqref{eq:rhogamma}. Here, one has $\kLo={\rm diag}(c_Q,c_Q)$ and $\kRo={\rm diag}(c_u,c_d)$. The first two terms in the above equation come from $\mathcal{L}_{\rm WZW}^{\rm full}$ and $\mathcal{L}_{\rm \chi PT}^{\rm ano}$, respectively, while the third term from the mass and kinetic mixings
between $a$ and $\pi_0$. Diagrammatically, the three contributions are depicted in Fig.~\ref{fig:Feynman-diagrams}.  

We emphasize the importance of including all three contributions to obtain the consistent results. Otherwise, one may obtain inconsistent couplings such as those in Ref.~\cite{Chakraborty:2024tyx} (see ~\footnote{Ref.~\cite{Chakraborty:2024tyx} (motivated by Ref.~\cite{Harvey:2007rd}) studied the constraints on the $a$-$\omega$-$\gamma$ interaction, hence the $c_{gg}$ coupling, from the cooling effects of astrophysical objects. Other than the mixing factors of $m_\pi^2/(m_a^2-m_\pi^2)$, their derivation of the $a$-$\omega$-$\gamma$ coupling missed the important contribution from the WZW term, which leads to a difference of about a factor of 2 compared to the result derived from the consistent treatment in this paper.}) and Ref.~\cite{Aloni:2018vki} (see ~\footnote{Ref.~\cite{Aloni:2018vki} adopts the interactions of pseudoscalar mesons with hidden local gauge bosons~\cite{Harada:2003jx} and then applies axion-meson mixings to obtain the axion interactions. While this method gives the correct $a$-$\gamma$-$\gamma$ coupling, it falls short of correctly producing the other interactions like $a$-$\omega$-$\gamma$ given in this work due to its lack of: (1) the anomalous interactions of $\mathcal{L}_{\chi {\rm PT}}^{\rm ano}$ from the quark chiral phase rotation or the change from $c_{\cA_1\cA_2}^0$ to $c_{\cA_1\cA_2}$ using the fictitious background gauge fields to track global anomalies, and (2) the full WZW term $\mathcal{L}_{\rm WZW}^{\rm full}$ that includes axion interactions (this term is absent for the $a$-$\gamma$-$\gamma$ coupling). As a result, the axion branching ratios predicted in their work are inaccurate and can be corrected by extending the framework of this work to the three-flavor scenario, which we leave to future study.} ). Other basis-independent axion couplings to the neutral electroweak gauge boson can be found in Eq.~\eqref{eq:a-Z-gamma} for $a$-$Z$-$\gamma$ and Eq.~\eqref{eq:a-Z-omega} for $a$-$Z$-$\omega$, which may have phenomenological consequences at neutrino experiments. 
\\

\begin{figure}[hbt]
\centering
\includegraphics[width= 0.99 \linewidth]{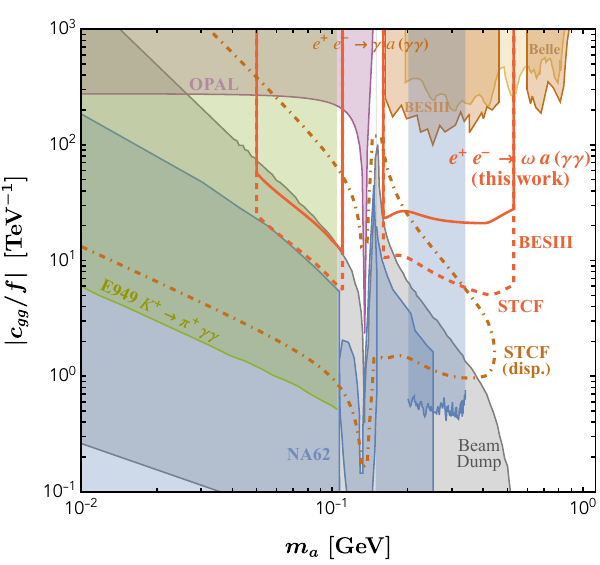}
\caption{The projected sensitivities to search for the axion particle via the process $e^+e^- \to \omega + a$ with a prompt decay of $a \rightarrow \gamma \gamma$ at the BESIII~\cite{BESIII:2021cxx} (red solid line) as well as at the future STCF collider~\cite{Achasov:2023gey} (red dashed line). The long-lived axion case with a displaced vertex is estimated for STCF (brown dot-dashed line). Also shown are other existing constraints including Beam Dump~\cite{CHARM:1985anb,Riordan:1987aw,Bjorken:1988as,Dolan:2017osp,Blumlein:1990ay,Dobrich:2019dxc,NA64:2020qwq}, NA62~\cite{NA62:2020pwi,NA62:2021zjw,NA62:2023olg}, E949~\cite{E949:2005qiy}, OPAL~\cite{Knapen:2016moh}, Belle II~\cite{Belle-II:2020jti}, and the BESIII search for $J/\psi \rightarrow \gamma a (\gamma\gamma)$~\cite{BESIII:2024hdv}. Except NA62 that directly probes the axion-gluon coupling, the constraints based on the $a$-$\gamma$-$\gamma$ coupling have been rescaled based on the corresponding axion diphoton decay branching ratio derived in this work.}
\label{fig:results} 
\end{figure} 

\noindent \textbf{New channels at low energy $e^+e^-$ colliders}.
The main focus of this paper is to derive the consistent axion interactions with various fundamental and composite gauge bosons including the WZW interactions. Some of the interactions derived here can already provide interesting phenomenological consequences. Here, we use the $a$-$\omega$-$\gamma$ coupling in Eq.~\eqref{eq:omegagamma} together with the  $a$-$\rho_0$-$\gamma$ coupling as an example to illustrate some novel collider signatures at low-energy $e^-e^+$ colliders (see Refs.~\cite{Bai:2023bbg,Chakraborty:2024tyx} for recent studies on the astrophysical consequences of this coupling). For simplicity, we consider the axion models with only nonzero $c_{gg}$ while $c_Q=c_u=c_d=0$. Since we only analyze the two-flavor chiral Lagrangian without the effects of axion mixings with $\eta$ or $\eta'$ in this work, we restrict the axion mass below around the $\eta$ mass or $m_a \le 500$~MeV. 

We first note that the coupling $c_{\rho\gamma}^{\rm eff}$ in Eq.~\eqref{eq:rhogamma} has a mild influence on the axion decay width and thus the branching ratios (see Fig.~\ref{fig:axionbr} in Appendix~\ref{app:axion-decay-width}). The main decay channel of axion is $a\rightarrow \gamma\gamma$, followed by $a \rightarrow 3 \pi$ and $a\rightarrow \rho_0^* \gamma \rightarrow \pi\pi\gamma$. For an axion mass lighter than the pion mass, the decay length of the axion is $c\tau_0 = 1.3\,\mbox{m}\times c_{gg}^{-2}\,(f/1\,\mbox{TeV})^2\,(0.1\,\mbox{GeV}/m_a)^3$. So, the lighter axion could be long-lived or even behave as a stable particle at colliders. 

At a low-energy $e^+e^-$ collider, one can search for the axion particle in the process of $e^+e^- \rightarrow \omega + a\,(a \rightarrow \gamma\gamma)$. For the prompt decay case, the $e^+e^-$ run with $\sqrt{s} = 3.097$~GeV ($J/\psi$ resonance mass) at the BESIII~\cite{BESIII:2021cxx} can probe some new parameter space in  $m_a$ and $|c_{gg}/f|$. In Appendix~\ref{app:pheno-ee-collider}, we perform a sensitivity study and show the projected limits in Fig.~\ref{fig:results} using the same amount of data in Ref.~\cite{BESIII:2021cxx} and including the form factor effects from vector meson dominance (see Appendix~\ref{app:omega-a-cross-section}). In Fig.~\ref{fig:results}, we also show the projected limits from the future
Super-Tau-Charm Facility (STCF)~\cite{Achasov:2023gey}, which has been proposed to run with a larger luminosity (10\,$\mbox{ab}^{-1}$) with good geometry and position resolutions to identify the long-lived axion case with a displaced vertex. For an axion stable at the collider length scale, it behaves as an invisible particle. In Appendix~\ref{app:omega+invisible}, we also estimate the sensitivity of $\omega$+invisible at STCF and find that the projected limits are superseded by the existing NA62 and Beam Dump limits, so we do not present the estimated limits in Fig.~\ref{fig:results}. In this figure, we do not show the constraints from $B^\pm \rightarrow K^\pm a (\gamma\gamma)$ from BaBar~\cite{BaBar:2021ich}, which depends on the UV cutoff scale and other UV flavor-changing couplings. 
\\

\noindent \textbf{Conclusions}. For two quark flavors, we have derived the complete axion-interaction Lagrangian below the QCD scale, encompassing the standard chiral Lagrangian, the WZW term, and the counter-term. This Lagrangian offers a comprehensive framework for understanding how axions interact with pseudo-scalar and (axial-)vector mesons, as well as the Standard Model gauge bosons. We have demonstrated that this Lagrangian successfully matches the effective Lagrangian at the quark level in two different bases, achieving consistent physical amplitudes for axions interacting with gauge bosons and vector mesons—an accomplishment not previously realized. Future work will extend this analysis to the three-quark-flavor case.
\\

\noindent \textbf{Acknowledgement}.
YB thanks Christopher T. Hill for illuminating discussion. JL thanks Haibo Li, Dayong Wang, Feng Xu and Shilin Zhu for useful discussions. The works of YB and TKC are supported by the U.S. Department of Energy under the contract DE-SC-0017647. The work of JL is supported by Natural Science Foundation of China under grant No. 12075005 and 12235001.

\bibliography{axionWZW}
\bibliographystyle{JHEP}

\onecolumngrid

\appendix

\begin{center}
{\Large Appendix for Wess-Zumino-Witten Interactions of Axions} \\[0.05 cm]
	\vspace{0.05 cm}
\end{center}

In this Appendix, we present the technical details supporting our manuscript for the convenience of our readers. The first section provides the complete formula for the effective Lagrangian at the quark level, along with the chiral Lagrangian incorporating the full Wess-Zumino-Witten interactions, and the explicit forms for the axion-associated Wess-Zumino-Witten interactions. In the second section, we demonstrate the matching correspondence between these two formulations. We also show that the comprehensive chiral Lagrangian inherently satisfies the vector meson dominance condition for the axion-gluon coupling case. We explicitly demonstrate that the physical vertices of axions remain invariant under auxiliary chiral phase rotations, ensuring consistent results. The third section contributes to the phenomenology of the model, including the calculations of axion decay widths and the specifics of the axion searches at low-energy electron-positron colliders.

\tableofcontents

\section{Axion Lagrangians across different energy scales}
Due to QCD confinement, the Lagrangians for axions differ across energy scales. Above the QCD scale but below the electroweak scale, we present an effective quark-level Lagrangian for axions. Below the QCD scale, we provide a comprehensive chiral Lagrangian that describes axion interactions. Notably, the inclusion of Wess-Zumino-Witten terms for axions results in a complete and consistent model in various respects.

\subsection{The effective quark-level Lagrangian for axions}
We begin with the effective axion Lagrangian applicable between the electroweak and QCD scales~\cite{Georgi:1986df}, which describes various axion couplings to the Standard Model quarks and vector gauge bosons, including gluons. Unlike previous literature, we explicitly include the anomalous couplings to vector mesons, for example $\omega$ and $\rho$. While it might seem confusing to include mesons above the QCD scale, we retain them to trace the mixed or pure anomalies associated with global symmetries. We will demonstrate later that these anomalies are indeed essential to the chiral Lagrangian at lower energies. 

We first introduce the SM electroweak gauge fields in terms of differential one-forms ($\mathbb{A}_{L,R}\equiv \mathbb{A}_{L,R}^\mu dx_\mu$)~\cite{Harvey:2007ca}
\begin{equation}
\begin{aligned}
    \mathbb{A}_L = \frac{e}{s_{\rm w}}W^a\frac{\boldsymbol{\tau}^a}{2} + \frac{e}{c_{\rm w}}W^0\mathbf{Y}_Q ,~ \mathbb{A}_R = \frac{e}{c_{\rm w}}W^0\mathbf{Y}_q ~,
\end{aligned}
\label{eq:gauge_fields:app}
\end{equation}
where $W^a$ and $W^0$ stands for the $SU(2)_L$ and $U(1)_Y$ hypercharge gauge one-forms, respectively, $\mathbf{Y}_Q={\rm diag}(1/6,1/6)$, and $\mathbf{Y}_q={\rm diag}(2/3,-1/3)$. Next, we write the fictitious background gauge boson field one-forms ($\mathbb{B}_{L,R}\equiv \mathbb{B}_{L,R}^\mu dx_\mu$) as
\begin{equation}
\begin{aligned}
    \mathbb{B}_V &\equiv \mathbb{B}_L + \mathbb{B}_R = g\begin{pmatrix}
        \rho_0 & \sqrt{2}\rho^+ \\
        \sqrt{2}\rho^- & -\rho_0
    \end{pmatrix} + g'\begin{pmatrix}
        \omega & \\
        & \omega
    \end{pmatrix} + (\kLo + \kRo)\frac{da}{\fa} ~, \\
    \mathbb{B}_A &\equiv \mathbb{B}_L - \mathbb{B}_R = g\begin{pmatrix}
        a_1 & \sqrt{2}a^+ \\
        \sqrt{2}a^- & -a_1
    \end{pmatrix} + g'\begin{pmatrix}
        f_1 & \\
        & f_1
    \end{pmatrix} + (\kLo - \kRo)\frac{da}{\fa} ~,
\end{aligned}
\label{eq:vec_meson_fields:app}
\end{equation}
where $\rho_0, \rho^{\pm}, \omega, a_1, a^{\pm}, f_1$ represent the isospin singlet or triplet (axial-)vector mesons. The definitions $V,A=(L\pm R)$ are slightly unconventional. Note that the axion fields are included in the background fields, which is essential for the consistency of the physical observables. We choose the phenomenological values $g=5.9,~g'=5.7$, based on the experimental analysis on $\omega\to \pi_0 \gamma$ and $\rho_0\to\pi\pi$~\cite{ParticleDataGroup:2022pth}. These values also agree quite well with the ``hidden local symmetry" approach in Ref.~\cite{Harada:2003jx} with $g'=g=5.8$.

The effective axion Lagrangian starts with the following form,
\begin{equation}
\label{eq:ALP:eff:before}
\begin{aligned}
    \mathcal{L}_{\rm eff,0} &= \mathcal{L}_{\rm SM} +  \bar{q}_0(i\slashed{D}-\mathbf{m}_{q,0})q_0 + \frac{1}{2}(\partial_\mu a)(\partial^\mu a) - \frac{m_{a,0}^2}{2}a^2  + c_{gg}\frac{\alpha_s}{4\pi}\frac{a}{\fa}\Gsub\Gtildesup + \frac{a}{\fa}\sum_{\cA_{1,2}} c_{\cA_{1} \cA_{2}}^0 F_{\cA_{1} \,\mu\nu}\,\widetilde{F}_{\cA_{2}}^{\mu\nu} +\mathcal{L}_c ~,
\end{aligned}
\end{equation}
where $\cA $ represents the $SU(2)_L\times U(1)_Y$ gauge bosons, $\mathbb{A}$, as well as the fictitious gauge bosons, $\mathbb{B}$, written in the flavor basis, and $\widetilde{G}^{\mu\nu} \equiv \frac{1}{2}\epsilon^{\mu\nu\alpha\beta}G_{\alpha\beta}$ and similarly for the other vector boson field strengths. The counter-term $\Gamma_c \equiv \int d^4 x\mathcal{L}_c$ is introduced to maintain the fundamental gauge invariance of the theory and is given by~\cite{Harvey:2007ca}
\begin{equation}\label{eq:WZW:counterterm}
\begin{aligned}
	\Gamma_c &= -2\mathcal{C}\int \Tr\Big[ (\mathbb{A}_Ld\mathbb{A}_L+d\mathbb{A}_L\mathbb{A}_L)\mathbb{B}_L + \frac{1}{2}\mathbb{A}_L(\mathbb{B}_Ld\mathbb{B}_L+d\mathbb{B}_L\mathbb{B}_L) - \frac{3}{2}i\mathbb{A}_L^3\mathbb{B}_L  - \frac{3}{4}i\mathbb{A}_L\mathbb{B}_L\mathbb{A}_L\mathbb{B}_L - \frac{i}{2}\mathbb{A}_L\mathbb{B}_L^3 \Big] - (L\leftrightarrow R) ~.
\end{aligned}
\end{equation}

To remain simple, we consider a two-flavor scenario with $q_0=(u,d)$ and $\mathbf{m}_{0}={\rm diag}(m_u, m_d)$, where $\kLo$ and $\kRo$ are Hermitian matrices representing the axion couplings to the left- and right-handed quarks, respectively. Extending this to three flavors is straightforward. The covariant derivative acting on the quark field is given by
\begin{equation}\label{eq:covariant:quark}
    D^\mu = \partial^\mu - i\sum_{\cA}g_{\cA}\left(\cA_{L}^\mu P_L + \cA_{R}^\mu P_R\right) ,\quad P_{L,R} = \frac{1\mp\gamma^5}{2} ~,
\end{equation}
where $g_{\cA}$ denotes the gauge couplings. It is worth noting that the charge-conjugation ($C$) symmetry is explicitly broken if $\kLo + \kRo \neq 0$.

Subsequently, we undertake a chiral rotation on the quark field to eliminate the $aG\tilde{G}$ interaction,~\cite{Srednicki:1985xd,Bardeen:1986yb,Georgi:1986df},
\begin{equation}
    q_0(x) = \exp\left[-i(\deltaqo+\kappaqo \gamma_5)c_{gg}\frac{a(x)}{\fa} \right]q(x) ,\quad \mbox{with}\; \Tr(\kappaqo)=1 ~,
\end{equation}
where $\deltaqo={\rm diag}(\delta_{u,0}, \delta_{d,0})$, $\kappaqo={\rm diag}(\kappa_{u,0}, \kappa_{d,0})$. We parameterize the rotation according to the chiral representation, $\boldsymbol{\theta}_{L/R,0}\equiv\deltaqo\mp\kappaqo ,~ \QLo = {\rm diag}(\theta_{Q,0}, \theta_{Q,0}) ,~ \QRo = {\rm diag}(\theta_{u,0}, \theta_{d,0})$. This will also effectively shift $\mathbf{m}_{q,0} \to \mathbf{m}_{q}(a)$ and  $\boldsymbol{k}_{L/R,0}\to\boldsymbol{k}_{L/R}(a)$. Subsequently, Eq.~\eqref{eq:ALP:eff:before} transforms into a new effective Lagrangian as:
\begin{equation} \label{eq:ALP:eff:after}
\begin{aligned}
    \mathcal{L}_{\rm eff} &= \mathcal{L}_{\rm SM} + \qbar\,i\slashed{D}\,q - [\qbar_L\,\mathbf{m}_q(a)\,q_R + h.c.] + \frac{1}{2}(\partial_\mu a)(\partial^\mu a) - \frac{m_{a,0}^2}{2}a^2 + \frac{a}{\fa}\sum_{\cA_{1,2}} c_{\cA_{1} \cA_{2}} F_{\cA_{1} \,\mu\nu}\,\widetilde{F}_{\cA_{2}}^{\mu\nu} + \mathcal{L}_c ~.
\end{aligned}
\end{equation}
Following this chiral rotation, the couplings exhibit the following relations~\cite{Bauer:2021wjo}:
\begin{equation}\label{eq:Leff:couplings}
\begin{aligned}
    \mathbf{m}_q(a) &= e^{ic_{gg}\QLo\, a/f}\,\mathbf{m}_{0}\,e^{-ic_{gg}\QRo\, a/f} ~, \\
    \kL(a) &= e^{ic_{gg}\QLo\,a/\fa}[\kLo+c_{gg}\QLo]e^{-ic_{gg}\QLo\,a/\fa} ~, \\
    \kR(a) &= e^{ic_{gg}\QRo\,a/\fa}[\kRo+c_{gg}\QRo]e^{-ic_{gg}\QRo\,a/\fa} ~, \\
    \frac{a}{\fa}\sum_{\cA_{1,2}} c_{\cA_{1} \cA_{2}} F_{\cA_{1} \,\mu\nu}\,\widetilde{F}_{\cA_{2}}^{\mu\nu} &= \frac{a}{\fa}\sum_{\cA_{1,2}} c_{\cA_{1} \cA_{2}}^0 F_{\cA_{1} \,\mu\nu}\,\widetilde{F}_{\cA_{2}}^{\mu\nu} - \delta\left[\mathcal{L}_{\rm WZW} + \mathcal{L}_c\right](\boldsymbol{c_{gg}\theta}_{L,0},c_{gg}\boldsymbol{\theta}_{R,0}) ~.
\end{aligned}
\end{equation}
For diagonal $\QLo$ and $\QRo$, one has $\kL(a) = \kLo+c_{gg}\QLo$ and $\kR(a) = \kRo+c_{gg}\QRo$ which are independent of $a$.  
Here, $\Gamma_{\rm WZW}\equiv \int d^4 x\mathcal{L}_{\rm WZW}$ and~\cite{Harvey:2007ca} 
\begin{align}
   \delta\left[\Gamma_{\rm WZW}+\Gamma_c\right](\boldsymbol{\theta}_{L},\boldsymbol{\theta}_{R}) &= -2\mathcal{C}\frac{a}{f}\int {\rm Tr}\Big\{\boldsymbol{\theta}_{L}\Big[3(d\mathbb{A}_L-i\mathbb{A}_L^2)^2+3(d\mathbb{A}_L-i\mathbb{A}_L^2)(D\mathbb{B}_L)+D\mathbb{B}_LD\mathbb{B}_L-\frac{i}{2}D(\mathbb{B}_L^3)\nonumber\\
   &+i\mathbb{B}_L(d\mathbb{A}_L-i\mathbb{A}_L^2)\mathbb{B}_L-i(d\mathbb{A}_L-i\mathbb{A}_L^2)\mathbb{B}_L^2 \Big]\Big\}~-~(L\leftrightarrow R) ~,
   \label{eq:anomalous}
\end{align}
where $D\mathbb{B}_{L,R} = d\mathbb{B}_{L,R} - i\mathbb{A}_{L,R}\mathbb{B}_{L,R} - i\mathbb{B}_{L,R}\mathbb{A}_{L,R}$, and $d\mathbb{A}_L-i\mathbb{A}_L^2$ is the covariant field strength. The first term of right-hand side is exactly the familiar axion-related Adler-Bell-Jackiw anomaly that plays an important role in other literature. We remark that the anomaly induced at this scale should match to that below the QCD scale by the WZW term, and thus the notation ``WZW''.
This constitutes a part of our consistent treatment of the axion field in following sections. Furthermore, as explained in the main text, the physical observables predicted by the effective Lagrangian $\mathcal{L}_{\rm eff}$ remain unaffected by the choice of auxiliary rotation parameters. For further insights, see for instance, Refs.~\cite{Bauer:2020jbp, Bauer:2021wjo}.

\subsection{The chiral Lagrangian for axions and the WZW interactions}

Below the QCD scale, the behavior of light-flavor QCD along with the axion can be described by the chiral Lagrangian together with the WZW term. While in principle, one can integrate out the heavy weak gauge bosons below the QCD scale, we retain them in the Lagrangian to illustrate the algebraic process. Thus, the chiral Lagrangian is expressed as follows:
\begin{equation} 
\label{eq:chiPT1}
\begin{aligned}
    \mathcal{L}_{\chi{\rm PT}} &= \frac{\fpi^2}{8}\Tr\left[(D^\mu U)(D_\mu U)^\dagger\right] + \frac{\fpi^2}{4}B_0\Tr\left[\mathbf{m}_q(a)U^\dagger + {\it h.c.}\right] + \frac{1}{2}(\partial_\mu a)(\partial^\mu a) - \frac{m_{a,0}^2}{2}a^2 + \frac{a}{\fa}\sum_{\cA_{1,2}} c_{\cA_{1} \cA_{2}} F_{\cA_{1} \,\mu\nu}\,\widetilde{F}_{\cA_{2}}^{\mu\nu} ~,
\end{aligned}
\end{equation}
where $U=\exp[(\sqrt{2}i/\fpi)\pi^a\boldsymbol{\tau}^a]$ describes the Goldstone fields of the $SU(2)_L\times SU(2)_R$ symmetry, $\boldsymbol{\tau}^a$ denoting the generator of $SU(2)_A$. Here, $\fpi\approx130$~MeV is the pion decay constant and $B_0\approx m_\pi^2/(m_u+m_d)$. We leave the counter-term $\mathcal{L}_c$ aside from the chiral Lagrangian $\mathcal{L}_{\chi{\rm PT}}$ and will include it together with the WZW term later.

Several points merit emphasis for Eq.~\eqref{eq:chiPT1}:
\begin{itemize}
\item Firstly, the anomalous axion coupling term $\mathcal{L}_{\chi{\rm PT}}^{\rm ano} \equiv \frac{a}{\fa}\sum_{\cA_{1,2}} c_{\cA_{1} \cA_{2}} F_{\cA_{1} \,\mu\nu}\,\widetilde{F}_{\cA_{2}}^{\mu\nu}$, derived from the chiral rotation to eliminate the axion-gluon coupling, must be incorporated into the chiral Lagrangian $\mathcal{L}_{\chi{\rm PT}}$. This reflects the anomaly matching condition that the anomalies associated with global symmetries (\eg\, baryon symmetry) are compensated for by the emergence of physical vector mesons (\eg\, $\omega$).
\item Secondly, we remark that the covariant derivative employed here incorporates the axion field, as indicated in Eq.~\eqref{eq:covariant:quark}.
\item Thirdly, both gauge bosons and fictitious gauge bosons are encompassed in the covariant derivative, where $\mathcal{A}_{L/R}$ run over the left- and right-handed chiral (fictitious) gauge bosons expressed in the flavor basis.
\end{itemize}

In terms of differential one-forms, the full WZW term, encompassing both the original WZW term and the associated counter-term needed to reproduce the quark-level anomalies~\cite{Harvey:2007ca}, is given by~\cite{Kaymakcalan:1983qq,Harvey:2007ca}
\begin{equation}\label{eq:WZW}
\begin{aligned}
	\Gamma_{\rm WZW}^{\rm full}(U,\cA_L,\cA_R) &= \int d^4x\mathcal{L}_{\rm WZW}^{\rm full} \\
        &= \Gamma_0(
 U) + \mathcal{C}\int\Tr\Big\{ (\cA_L\alpha^3+\cA_R\beta^3) - \frac{i}{2}[(\cA_L\alpha)^2-(\cA_R\beta)^2] \\
	&\quad +i(\cA_LU\cA_R^\dagger\alpha^2-\cA_RU^\dagger\cA_LU\beta^2) + i(d\cA_RdU^\dagger\cA_LU-d\cA_LdU\cA_RU^\dagger) \\
	&\quad + i[(d\cA_L\cA_L+\cA_Ld\cA_L)\alpha + (d\cA_R\cA_R+\cA_Rd\cA_R)\beta] \\
	&\quad + (\cA_L^3\alpha+\cA_R^3\beta) - (d\cA_L\cA_L+\cA_Ld\cA_L)U\cA_RU^\dagger \\
	&\quad + (d\cA_R\cA_R+\cA_Rd\cA_R)U^\dagger\cA_LU + (\cA_LU\cA_RU^\dagger\cA_L\alpha + \cA_RU^\dagger\cA_LU\cA_R\beta) \\
	&\quad + i\Big[\cA_L^3U\cA_RU^\dagger - \cA_R^3U^\dagger\cA_LU - \frac{1}{2}(U\cA_RU^\dagger\cA_L)^2\Big] \Big\} + \Gamma_c ~,
\end{aligned}
\end{equation}
where $\Gamma_c$ is given in Eq.~\eqref{eq:WZW:counterterm}, $\mathcal{C}=N_c/48\pi^2$~\footnote{After careful examination, we find that due to convention matters, we should add an overall minus sign to $\Gamma_{\rm WZW}+\Gamma_c$ compared to Ref.~\cite{Harvey:2007ca} (which we denote as $\Gamma_{\rm WZW}^{\rm HHH}+\Gamma_c^{\rm HHH}$) in order to reproduce the global anomalies, and this convention is aligned with Ref.~\cite{Kaymakcalan:1983qq} . This special overall minus sign is also mentioned in reference 31 of Ref.~\cite{Harvey:2007ca}. In the following we define $\Gamma_{\rm WZW/c}=-\Gamma_{\rm WZW/c}^{\rm HHH}$ by setting $\mathcal{C}=N_c/48\pi^2$}, $\cA_{L,R}=\mathbb{A}_{L,R}+\mathbb{B}_{L,R}$, $\alpha=dUU^\dagger$, $\beta=U^\dagger dU$, and
\begin{equation}
	\Gamma_0(U) = -\frac{i\mathcal{C}}{5}\int d^5x\epsilon^{ABCDE}\Tr(\alpha_A\alpha_B\alpha_C\alpha_D\alpha_E) ~.
\end{equation}

It is crucial to emphasize that the axion one-form is introduced as a fictitious background gauge boson field corresponding to the global $U(1)_{\rm PQ}$ symmetry, as we have indicated in Eq.~\eqref{eq:vec_meson_fields:app}.
Therefore, we obtain the comprehensive chiral Lagrangian for the axions, ensuring the consistency of physical results. The final full axion interaction Lagrangian is presented as:
\begin{align}
    \mathcal{L}_{\rm axion}^{\rm full} \left(U, \mathbf{m}_q(a), \cA_{L/R} + \mathbf{k}_{L/R}(a) da/f\right)
    = \left[\mathcal{L}_{\chi{\rm PT}}+\mathcal{L}_{\rm WZW}^{\rm full} \right] \left(U, \mathbf{m}_q(a), \cA_{L/R} + \mathbf{k}_{L/R}(a) da/f\right).
\end{align}

\subsection{The axion-associated terms in WZW interactions}
\label{sec:axion-WZW-terms}

In this subsection, we explicitly list the axion-relevant terms from the WZW interactions, $\mathcal{L}_{\rm WZW}^{\rm full}$, for the convenience of readers. The abbreviated notation for differential forms is $\int d^4x \epsilon_{\mu\nu\rho\sigma}A^{\mu}B^{\nu}\partial^{\rho}C^{\sigma}\equiv \int ABdC$, as aligned with Ref.~\cite{Harvey:2007ca}. We parametrize explicitly $\kL={\rm diag}(k_Q,k_Q)$ and $\kR={\rm diag}(k_u,k_d)$. The three-point vertex terms in the form of $XdYda$, where $X$ and $Y$ represent gauge fields or vector meson fields, are given by
\begin{align}
\Gamma_{XdYda}&=\frac{\mathcal{C}}{f}\int da\left\{ \frac{2e^2}{s_{2{\rm w}}}(k_d+2k_u+3k_Q)\gamma dZ+eg(k_d+2k_u+3k_Q)\gamma da_1-eg'(k_d-k_Q-2k_u)\gamma df_1\right.\nonumber\\
&\left.+eg(k_d-3k_Q+2k_u)\gamma d\rho_0-eg'(k_d+k_Q-2k_u)\gamma d\omega+\frac{2e^2}{s_{2{\rm w}}^2}\left[(k_d+4k_Q+k_u)-2s_{\rm w}^2(k_d+3k_Q+2k_u)\right]ZdZ\right.\nonumber\\
&\left.+\frac{eg}{s_{2{\rm w}}}\left[(k_d+4k_Q+k_u)-2s_{\rm w}^2(k_d+3k_Q+2k_u)\right]Zda_1-\frac{eg'}{s_{2{\rm w}}}\left[k_d-k_u+s_{\rm w}^2(-2k_d+2k_Q+4k_u)\right]Zdf_1\right.\nonumber\\
&\left.-\frac{eg}{s_{2{\rm w}}}\left[-3k_d-3k_u+2s_{\rm w}^2(k_d-3k_Q+2k_u)\right]Zd\rho_0-\frac{eg'}{s_{2{\rm w}}}\left[3k_d-3k_u-2s_{\rm w}^2(k_d+k_Q-2k_u)\right]Zd\omega\right.\nonumber\\
&\left.+g^2(k_d+2k_Q+k_u)a_1d\rho_0+gg'(k_u-k_d)a_1d\omega+gg'(k_u-k_d)f_1d\rho_0+g'^2(k_d+2k_Q+k_u)f_1d\omega\right.\nonumber\\
&\left.+g^2(k_d-2k_Q+k_u)\rho_0 d\rho_0+2gg'(k_u-k_d)\rho_0d\omega+g'^2(k_d-2k_Q+k_u)\omega d\omega+\frac{3eg}{2s_{\rm w}}(k_u+k_d)W^\pm d\rho^\mp\right.\nonumber\\
&\left.+\frac{eg}{2s_{\rm w}}(k_d+4k_Q+k_u)a^\mp dW^\pm +g^2(k_d+2k_Q+k_u)a^\mp d\rho^\pm+\frac{e^2}{s_{\rm w}^2}(k_d+4k_Q+k_u)W^-dW^+\right\} , 
\end{align}
with the overall factor $\mathcal{C}=N_c/48\pi^2$. Similarly, the four-point vertex terms in the form of $XYZda$, are given by
\begin{align}
\Gamma_{XYZda}&=\pm i\frac{\mathcal{C}}{f}\int da\left\{ eg^2(k_d+2k_Q+k_u)\gamma a^\mp\rho^\pm+\frac{e^2g}{2s_{\rm w}}(k_d+4k_Q+k_u)\gamma a^\mp W^\pm-\frac{g^3}{4}(k_d+2k_Q+k_u)a^\mp a_1W^\pm\right.\nonumber\\
&\left.+\frac{eg^2}{s_{2{\rm w}}}(1-2s_{\rm w}^2)(k_d+2k_Q+k_u)Za^\mp\rho^\pm+\frac{e^2gc_{\rm w}}{s_{\rm w}^2}(k_d+4k_Q+k_u)Za^\mp W^\pm+\frac{3g^3}{4}(k_d+2k_Q+k_u)a^\mp\rho^\pm\rho^0\right.\nonumber\\
&\left.+\frac{e^2g}{s_{2{\rm w}}s_{\rm w}}\left[2k_d+2k_Q+2k_u-3s_{\rm w}^2(k_u+k_d)\right]Z\rho^\mp W^\pm-\frac{g^3}{4}(k_d-2k_Q+k_u)a^\mp a_1\rho^\pm-\frac{g^2g'}{4}(k_u-k_d)a^\mp\rho^\pm f_1\right.\nonumber\\
&\left.-\frac{3g^2g'}{4}(k_u-k_d)a^\mp\rho^\pm\omega-\frac{g^2e}{2s_{\rm w}}(k_d+2k_Q+k_u)a^\mp\rho_0W^\pm-\frac{g^3}{4}(k_d-2k_Q+k_u)a_1a^\pm\rho^\mp+\frac{g^2e}{2s_{\rm w}}(k_d+2k_Q+k_u)a_1\rho^\mp W^\pm\right.\nonumber\\
&\left.-\frac{g^2e}{2s_{\rm w}}(3k_d-2k_Q+3k_u)\rho^\mp\rho^0W^\pm-\frac{gg'e}{s_{\rm w}}(k_u-k_d)\rho^\mp f_1W^\pm+\frac{g^2e}{2s_{\rm w}}(3k_d-2k_Q+3k_u)\rho^\pm\rho^0W^\mp\right\}\nonumber\\
&+i\frac{\mathcal{C}}{f}\int da\left\{\frac{eg^2}{s_{2{\rm w}}}\left[3k_d-2k_Q+3k_u-4s_{\rm w}^2(k_d-2k_Q+k_u)\right]Z\rho^-\rho^++\frac{eg^2}{s_{2{\rm w}}}(k_d+2k_Q+k_u)Za^-a^+ 
\right.\nonumber\\
&\left. +2eg^2(k_u+k_d-2k_Q)\gamma \rho^-\rho^+ + \frac{e^3}{2s_w^2s_{2{\rm w}}}\left[-3k_d-18k_Q-3k_u+4s_{\rm w}^2(k_d+4k_Q+k_u)\right]ZW^+W^-
\right.\nonumber\\
&\left.-\frac{3g^3}{4}(k_d+2k_Q+k_u)a^-a_1a^++\frac{g^3}{4}(k_d-2k_Q+k_u)a^-a^+\rho_0+\frac{g^2g'}{4}(k_u-k_d)a^-a^+f_1+\frac{3g^2g'}{4}(k_u-k_d)a^-a^+\omega
\right.\nonumber\\
&\left.+\frac{3g^3}{4}(k_d+2k_Q+k_u)a_1\rho^-\rho^+-\frac{ge^2}{2s_{\rm w}^2}(k_d+4k_Q+k_u)a_1W^+W^-+\frac{9g^3}{4}(k_d-2k_Q+k_u)\rho^-\rho^+\rho_0
\right.\nonumber\\
&\left.+\frac{9g^2g'}{4}(k_u-k_d)\rho^-\rho^+f_1+\frac{3g^2g'}{4}(k_u-k_d)\rho^-\rho^+\omega
-\frac{ge^2}{s_{\rm w}^2}(k_d+k_Q+k_u)\rho_0W^+W^-
-\frac{g'e^2}{2s_{\rm w}^2}(k_u-k_d)f_1W^+W^-\right\} ~.
\end{align}

We remark that the $C$-violating interactions, such as $da \, \gamma \, da_1$, arise from the possible $C$-violating axion quark-bilinear couplings mentioned earlier, {\it i.e.,} $\kLo+\kRo\neq0$.

\section{The consistent matching between the quark-level and chiral Lagrangians}

The consistent matching between the effective quark-level Lagrangian and the chiral Lagrangian for pseudoscalar mesons has been demonstrated in Ref.~\cite{Bauer:2021wjo}. In the previous section, we extended the chiral Lagrangian to include vector mesons and the WZW interactions for axions. We will show that the matching of the two Lagrangians in one chiral basis automatically ensures consistent matching in the other chiral basis.
Additionally, the full chiral Lagrangian can be used to derive consistent physical vertices for axion-meson interactions, which remain invariant under auxiliary chiral rotations. In this section, we will enumerate the important ones and provide the full results of these vertices.

\subsection{The axion-dependent chiral basis transformation}
While one is allowed to choose any chiral basis to perform the calculation, the physical results should be independent of the basis choice.
Above the QCD scale, the axion-dependent chiral basis transformation is given by
\beqa
q^\prime =  \exp \Bigl[  i\left(\boldsymbol{\delta}_q +\boldsymbol{\kappa}_q  \gamma_5 \right) a/f \Bigr]  q ~,
\label{eq:auxiliary-chiral-rotation}
\eeqa
with $\boldsymbol{\theta}_{L/R} \equiv \boldsymbol{\delta}_q \mp  \boldsymbol{\kappa}_q$ and $U_{L/R} \equiv  \exp \bigl[  -i \boldsymbol{\theta}_{L/R}   a/f \bigr]$. 
$\boldsymbol{\theta}_{L/R}$ are in general Hermitian matrices chosen to be diagonal in the quark mass basis.  Therefore, the quark mass and quark bi-linear couplings transform as
\begin{align}
\mathbf{m}'_q = U_L^\dagger \mathbf{m}_q U_R , \quad 
\mathbf{k}'_{L/R}= U_{L/R}^\dagger(\mathbf{k}_{L/R} + \boldsymbol{\theta}_{L/R} )U_{L/R}=\mathbf{k}_{L/R} + \boldsymbol{\theta}_{L/R}~.
\end{align}
Note that we have already started from the chiral Lagrangian with the axion-gluon interaction eliminated. Therefore, the chiral transformation should satisfy $\Tr(\boldsymbol{\kappa}_q) = 0$ to avoid regenerating the axion-gluon interaction.
From the $q$ basis to the $q'$ basis, the effective Lagrangian is transformed to be 
\beqa
\mathcal{L}_{\rm eff}(q, \mathbf{m}_q, \mathbf{k}_L,\mathbf{k}_R) \rightarrow \mathcal{L}_{\rm eff}(q', \mathbf{m}'_q, \mathbf{k}'_L,\mathbf{k}'_R) + \delta \mathcal{L}^{\rm ano}_{a} ~,
\label{eq:Leffchange-1}
\eeqa
with $\delta \mathcal{L}^{\rm ano}_{a}$ tracking the additional axion anomalous interactions,
\begin{align}
    \delta \mathcal{L}^{\rm ano}_{a} = - \delta\left[\mathcal{L}_{\rm WZW} + \mathcal{L}_c\right](\boldsymbol{\theta}_{L},\boldsymbol{\theta}_{R}) ~,
\end{align}
with the formula given in Eq.~\eqref{eq:anomalous}.

On the other hand, below the QCD scale, one is also allowed to perform an auxiliary axion-dependent basis transformation. Under the left/right-handed chiral rotations of quarks, the pseudoscalar meson field $U=\exp[(\sqrt{2}i/\fpi)\pi^a\boldsymbol{\tau}^a]$ transforms as~\cite{Bauer:2021wjo,Kaymakcalan:1983qq,Harvey:2007rd}
\begin{align}
    U =  U_L U' U_R^\dagger \, .
\end{align}
Therefore, the kinetic and mass terms of the chiral Lagrangian change accordingly,
\begin{equation}
    \mathcal{L}_{\chi{\rm PT}} \supset \frac{f_\pi^2}{8}\Tr\left[D'^{\mu} U'\left(D_\mu' U'\right)^\dagger\right]+\frac{f_\pi^2}{4}B_0\Tr \left[\mathbf{m}_q^{\prime} U'^\dagger+{\it h.c.}
\right] ~,
\end{equation}
and the covariant derivative transforms as
\begin{equation}
iD_{\mu}'U'=i\partial_{\mu}U'+(\mathcal{A}_L)_{\mu}U'-U'(\mathcal{A}_R)_{\mu}+\frac{\partial_\mu a}{f}\left[\kL' U'-U' \kR' \right] ~.
\end{equation} 
It is straightforward to show that the basis transformation simply adds a prime to all the terms in the chiral Lagrangian, similar to Eq.~\eqref{eq:Leffchange-1}.

Now for the WZW term and its counter-term. After substituting $U$ with $U_L U' U_R^\dagger$, one has 
\begin{align}
 \Gamma_{\rm WZW}\left( U_L U' U_R^\dagger, \mathcal{A}_L, \mathcal{A}_R \right)+\Gamma_c \left(\mathbb{A}_{L/R}, \mathbb{B}_{L/R}\right) ~.
\label{eq:before_trans}
\end{align}
Note that $\mathcal{A}_{L/R}\equiv \mathbb{A}_{L/R}+\mathbb{B}_{L/R}$ and $\mathbf{k}_{L/R}\frac{da}{f}\subset \mathbb{B}_{L/R}$. We quote the transformation relation from Ref.~\cite{Harvey:2007ca} 
\begin{align}
     &\Gamma_{\rm WZW}\left( U = U_L U' U_R^\dagger, \mathcal{A}_L, \mathcal{A}_R \right)+\Gamma_c \left(\mathbb{A}_{L/R}, \mathbb{B}_{L/R}\right)\nonumber\\
     &= \Gamma_{\rm WZW}\left(U', \mathcal{A}'_L, \mathcal{A}'_R \right)+\Gamma_c \left(\mathbb{A}_{L/R}, \mathbb{B}'_{L/R}\right)-\delta\left[\Gamma_{\rm WZW}+\Gamma_c\right](\boldsymbol{\theta}_{L},\boldsymbol{\theta}_{R})\nonumber \\
     &\equiv \Gamma_{\rm WZW}\left(U', \mathcal{A}'_L, \mathcal{A}'_R \right)+\Gamma_c \left(\mathbb{A}_{L/R}, \mathbb{B}'_{L/R}\right) + \int d^4x\,\delta \mathcal{L}^{\rm ano}_{\rm WZW} ~,
     \label{eq:after_trans}
\end{align}
where, under axion-dependent chiral transformation, one has
\begin{equation}
      U= U_L U' U_R^\dagger\to U',~~ \delta \mathbb{A}_{L/R}=0,~~\delta \mathbb{B}_{L/R     }\equiv\mathbb{B}'_{L/R}-\mathbb{B}_{L/R}=\boldsymbol{\theta}_{L/R}da/f ~.
\end{equation}
Note that the last transformation is equivalent to shifting
$\mathbf{k}_{L/R}\to\mathbf{k}'_{L/R}$.

This equivalence between the anomalous term from a basis change in the quark level and that in the meson level, or equivalently $\delta\mathcal{L}^{\rm ano}_a = \delta \mathcal{L}^{\rm ano}_{\rm WZW}$, demonstrates the consistency of the matching procedure.

\subsection{The consistent physical amplitudes}
\label{app:physical-amplitude}

In this subsection, we demonstrate that the chiral axion Lagrangian yields consistent physical amplitudes that remain invariant under the auxiliary chiral rotation in Eq.~\eqref{eq:auxiliary-chiral-rotation}. First, we will provide the physical amplitude for \(a \, \gamma \, \gamma\) as an introductory example. Then, we will consider \(a \, X_R \, X_R\), where \(X_R\) is a fictitious chiral vector gauge boson coupling exclusively to right-handed fermions. Next, we will examine \(a \, \omega \, \gamma\) for the case involving both gauge and background vector fields, followed by \(a \, \gamma \, Z\), with \(Z\) representing the neutrino or lepton currents at low energy. Lastly, we will analyze \(a \, \omega \, Z\) for another instance of a mixed coupling among gauge and background vector fields.

It is inspiring to trace in detail how the dependence on these auxiliary rotation parameters cancels out in the physical amplitudes. We will explicitly demonstrate the cancellation process for several cases and provide the physical amplitudes for the aforementioned vertices for the readers' reference and convenience. In our simplified two-flavor scheme, this involves the non-trivial axion mass and kinetic mixing with \(\pi_0\), as well as subtle contributions from the WZW term and the counter-term.

\subsubsection{Axion-pion mixing}

We first derive the kinetic and mass mixings between \(a\) and \(\pi_0\), which are crucial for obtaining consistent physical amplitudes.
Given the Lagrangian in Eq.~\eqref{eq:chiPT1}, we expand the kinetic terms and find the kinetic mixing term as
\begin{align}
&\frac{f_\pi^2}{8}\left[i\frac{\partial_\mu a}{f}\Tr[\partial^\mu U(\kL(a)-\kR(a))]+{\it h.c.}\right]
=\frac{f_\pi^2}{8}\left[i\frac{\partial_\mu a}{f}i\sqrt{2}\frac{\partial^\mu \pi_\alpha}{f_\pi}\Tr[\tau_\alpha(\kL(a)-\kR(a))]+{\it h.c.}\right]+\mathcal{O}(1/f_\pi^2) , \nonumber\\
&\supset-\frac{\sqrt{2}f_\pi}{4f}\partial_\mu a\partial^\mu \pi_0\Tr[\tau_3(\kL(a)-\kR(a))]~~\text{(kinetic mixing)} ~,
\end{align}
and the mass mixing term as
\begin{align}
&-\frac{\sqrt{2}f_\pi}{f}\frac{m_\pi^2}{m_u+m_d}\Tr[\mathbf{m}_{q,0}\,\kappaqo\,\tau_3]\,a\,\pi_0~~\text{(mass mixing)} ~.
\end{align}
Therefore, we obtain the mixing angle between the $a$ and $\pi_0$ fields
\begin{align} 
   \theta_{\pi a}&=\frac{f_\pi}{2\sqrt{2}f}\left[\frac{m_a^2}{m_a^2-m_\pi^2}\Tr[\tau_3(\kL(a)-\kR(a))]-\frac{m_\pi^2}{m_\pi^2-m_a^2}\delta_{\kappa,0}\right] ~,\nonumber\\
   \pi_0&=\pi_0^{\rm phys}+\theta_{\pi a} ~a^{\rm phys} ~,
\end{align}
to the leading order in $(f_\pi/f)$, where $\displaystyle\dkappao \equiv 4c_{gg}(m_u\kappa_{u,0} - m_d\kappa_{d,0})/(m_u+m_d)$.

Following Ref.~\cite{Bauer:2021wjo}, we drop the subscript ``phys'' and expand the non-anomalous part of $\mathcal{L}_{\chi{\rm PT}}$ to the leading order, keeping only the terms related to the axion,
\begin{equation}
\begin{aligned}
    \mathcal{L}_{\chi{\rm PT}}^{\rm LO,a} &= \frac{1}{2}(\partial_\mu a)(\partial^\mu a) - \frac{m_a^2}{2}a^2 + \frac{1}{2}\partial_\mu\pi^0\partial^\mu\pi^0 + D_\mu\pi^+D^\mu\pi^- - \frac{m_\pi^2}{2}(\pi^0)^2 - m_\pi^2\pi^+\pi^- \\
    &\quad - \frac{\deltacud}{6\sqrt{2}\fpi\fa}\frac{m_\pi^2}{m_a^2-m_\pi^2}\Big[ 4\partial^\mu a(\pi^0\pi^+D_\mu\pi^- + \pi^0\pi^-D_\mu\pi^+ - 2\pi^+\pi^-\partial_\mu\pi^0) + m_\pi^2\,a\,(2\pi^+\pi^-\pi^0+(\pi^0)^3)\Big] ~,
\end{aligned}
\end{equation}
with 
 \begin{equation}
     \deltacud = (c_{u} - c_{d}) + 2c_{gg}\frac{m_d-m_u}{m_d+m_u},~~~~\kRo - \kLo = {\rm diag}(c_{u}-c_Q, c_{d}-c_Q) ~.\nonumber
\end{equation}

We note that the physical matrix element of a typical axion scattering process consists of two parts: one from the direct interaction between the axion and other particles, and the other from the axion mixing with \(\pi^0\). 
It has been shown that such processes should be indeed \(\kappa\)-independent, for example $a \to \gamma \gamma$~\cite{Bauer:2020jbp} and $K \to \pi a$~\cite{Bauer:2021wjo}. 
After imposing the \(\Tr(\kappa_{q, 0}) = 1\) constraint, the remaining \(\kappa\)-dependence in the axion and \(\pi^0\) components of the matrix element cancels out exactly. In fact, as we will show later, as long as \(\Tr(\kappa_{q, 0})\) is constrained to be a constant value, the physical matrix element will always be \(\kappa\)-independent at the leading order. 


\subsubsection{$a$-$\gamma$-$\gamma$ }
\label{app:a-gamma-gamma}

The axion-photon-photon scattering amplitude is fundamental and significant, as many axion searches are based on this interaction. It has already been discussed and consistently calculated using the effective quark-level Lagrangian \(\mathcal{L}_{\rm eff}\) in Ref.~\cite{Bauer:2020jbp}. However, it is delightful to revisit this process within the framework of the chiral axion Lagrangian to examine how the dependence on the auxiliary rotation parameters cancels out in the presence of the WZW term and the counter-term. 

The physical amplitude receives contributions from both the kinetic and mass mixing of \(a-\pi^0\), as well as from the interactions described by the full WZW interactions \(\mathcal{L}_{\rm WZW}^{\rm full}\) and the anomaly contribution \(\delta \mathcal{L}_{\rm WZW}^{\rm ano}\). For vector-like fundamental gauge fields, such as the SM photon field \(\gamma\), there is no contribution from \(\mathcal{L}_{\rm WZW}^{\rm full}\).
Therefore, under the basis transformation in Eq.~\eqref{eq:auxiliary-chiral-rotation}, we trace the auxiliary rotation parameters in each of these contributions:
\begin{itemize}
\item[$\bullet$] From anomaly contribution \(\delta \mathcal{L}_{\rm WZW}^{\rm ano}\)
\begin{align}
ad\gamma d\gamma:~~c_{\rm ano} \equiv -\frac{e^2N_c}{48\pi^2f}12(Q_u^2\kappa_u+Q_d^2\kappa_d) ~,
\end{align}
\item[$\bullet$] From the $\pi_0$ contribution 
\begin{align}
\pi_0d\gamma d\gamma:~~c_{\pi_0} \equiv \frac{e^2N_c}{48\pi^2f_\pi}6\sqrt{2}(Q_d^2-Q_u^2) ~,
\end{align}
\item[$\bullet$] From the full WZW interactions \(\mathcal{L}_{\rm WZW}^{\rm full}\)
\begin{align}
ad\gamma d\gamma:~~c_{\rm wzw} \equiv 0~~~~\text{(vector-like fundamental gauge field)} ~.
\end{align}
\end{itemize}
The counter-term ensures that for vector-like gauge fields, \textit{i.e.}, $\mathbb{A}_L=\mathbb{A}_R=\mathbb{A}$, the  pseudo-Chern-Simons terms like $da\mathbb{A}d\mathbb{A}$ are totally canceled out~\cite{Harvey:2007ca}. We define the common factor
\begin{align} 
CF \equiv \left<\gamma\gamma|a\,d\gamma d\gamma|a\right> ~,
\end{align}
in the amplitude to account for $a d\gamma d\gamma $ and combine the auxiliary rotation parameters in the three pieces together to obtain
\begin{align}
&\mathcal{M}(a\to\gamma\gamma)\text{(auxiliary)}=CF \times \left( c_{\rm ano} + \theta'_{a-\pi_0} c_{\pi_0} + c_{\rm WZW} \right) \\
 &= CF \times e^2\left\{\frac{-N_c}{48\pi^2f_a}12(Q_u^2\kappa_u+Q_d^2\kappa_d)+i\frac{f_\pi}{\sqrt{2}f}\left[(\kappa_u-\kappa_d)p_a^2-2\frac{m_u\kappa_u-m_d\kappa_d}{m_u+m_d}m_\pi^2\right]\frac{i}{p_a^2-m_\pi^2}\right.\nonumber\\
&\left.\times\frac{N_c}{48\pi^2f_\pi}6\sqrt{2}(Q_d^2-Q_u^2)\right\} ~.
\end{align}
$ p_a $ is the momentum of the axion. The $ p_a^2 $ term arises from the axion-pion kinetic mixing with $p_a^2 = m_a^2$ for an on-shell axion, while the term proportional to  \( m_\pi^2 \) originates from the axion-pion mass mixing. The axion-pion mixing angle component containing only the auxiliary parameters $\kappa_{u/d},~\delta_{u/d}$ is indicated in the second line with
\begin{align}
    \theta_{a-\pi_0}' = -\frac{f_\pi}{\sqrt{2}f} \left[(\kappa_u-\kappa_d)\frac{m_a^2}{m_a^2-m_\pi^2}-2\frac{m_u\kappa_u-m_d\kappa_d}{m_u+m_d}\frac{m_\pi^2}{m_a^2-m_\pi^2} \right] ~.
\end{align}

By reorganizing the two auxiliary rotation parameters into new linear combinations,
\beqa
    Q_u^2\kappa_u+Q_d^2\kappa_d &=&(Q_u^2-Q_d^2)(\kappa_u-\kappa_d)/2+(Q_u^2+Q_d^2)(\kappa_u+\kappa_d)/2 ~, \nonumber \\
\frac{m_u\kappa_u-m_d\kappa_d}{m_u+m_d} &=& \frac{\kappa_u - \kappa_d}{2} + \frac{m_u - m_d}{m_u + m_d} \frac{\kappa_u + \kappa_d}{2}  ~,
    \label{eq:linearsplit}
\eeqa
and imposing the constraint $\kappa_u+\kappa_d=0$ to avoid regenerating the $a G \tilde G$ term, we can easily verify that the auxiliary rotation parameters in the amplitude cancel out, leading to $\mathcal{M}(a\to\gamma\gamma)\text{(auxiliary)} = 0$. 

Finally, we write down the consistent physical result for the $a$-$\gamma$-$\gamma$ scattering amplitude, incorporating the effect arising from rotating out $aG\tilde{G}$
\begin{align}
& \mathcal{M}(a\to\gamma\gamma) = c_{\gamma \gamma}^{\rm eff} \times CF ~, \\
&c_{\gamma \gamma}^{\rm eff} = c_{\gamma\gamma}^0+c_{\rm ano}\frac{Q_u^2+Q_d^2}{2(Q_u^2\kappa_{u,0}+Q_d^2\kappa_{d,0})}-c_{\pi_0}\frac{f_\pi}{\sqrt{2}f}\left(\frac{m_a^2}{m_\pi^2-m_a^2}\frac{\Tr[\tau_3(\kLo-\kRo)]}{2}-c_{gg}\frac{m_u-m_d}{m_u+m_d}\frac{m_\pi^2}{m_a^2-m_\pi^2}\right) \\
&=c_{\gamma\gamma}^0+e^2\left\{\frac{-N_cc_{gg}}{48\pi^2f}6(Q_u^2+Q_d^2)+\frac{1}{\sqrt{2}f}\left[\frac{m_a^2}{m_\pi^2-m_a^2}\left(\frac{\Tr[\tau_3(\kLo-\kRo)]}{-2}\right)+c_{gg}\frac{m_u-m_d}{m_u+m_d}\frac{m_\pi^2}{m_a^2-m_\pi^2}\right]\times \frac{N_c}{48\pi^2}6\sqrt{2}(Q_d^2-Q_u^2)\right\} ~, \nonumber 
\end{align}
where $c_{\gamma\gamma}^0$ comes from the contribution at the UV scale. Substituting $N_c=3$, $Q_u=2/3$ and $Q_d=-1/3$, we arrive at the final result
\begin{align}
c_{\gamma\gamma}^{\rm eff}=c_{\gamma\gamma}^0+\frac{e^2c_{gg}}{16\pi^2 f}\left(-\frac{10}{3}-2\frac{m_u-m_d}{m_u+m_d}\frac{m_\pi^2}{m_a^2-m_\pi^2}\right)-\frac{e^2}{16\pi^2 f}\frac{m_a^2}{m_\pi^2-m_a^2} (c_u-c_d) ~.
\label{eq:agammagamma}
\end{align}
This result is consistent with Eq.~(92) of Ref.~\cite{Bauer:2020jbp}. Note that in our notation, $dAdA = \frac{1}{2}F^{\mu\nu}\widetilde{F}_{\mu\nu}~d^4x$. It is worth pointing out that since the fundamental gauge field \(\gamma\) is vector-like, we have \(c_{\rm WZW} = 0\), making the WZW and its counter-term interactions trivial in this process. Next, we will show that the WZW and its counter-term interactions are important for chiral fundamental gauge bosons. Including the full WZW term is essential for a consistent physical amplitude.

\subsubsection{$a$-$\gamma_R$-$\gamma_R$ }

To demonstrate the importance of incorporating the axion into the full WZW term, we propose a chiral fundamental gauge boson \(\gamma_R\) for a \(U(1)'\) Abelian gauge group as a toy model, which couples exclusively to right-handed fermions and not to left-handed fermions at all. The axion field in this model also only couples to the right-handed fermions. Similar to the previous subsection, under the basis transformation in Eq.~\eqref{eq:auxiliary-chiral-rotation}, we trace the auxiliary rotation parameters in each of these contributions, using the notations \(\alpha_{u/d} = \delta_{u/d} + \kappa_{u/d}\):
\begin{itemize}
\item[$\bullet$] From anomaly contribution \(\delta \mathcal{L}_{\rm WZW}^{\rm ano}\)
\begin{align}
ad\gamma_R d\gamma_R:~~c_{\rm ano}=\frac{-e^2N_c}{48\pi^2f}6(Q_u^2\alpha_u+Q_d^2\alpha_d)~,
\end{align}
\item[$\bullet$] From the $\pi_0$ contribution 
\begin{align}
\pi_0 d\gamma_R d\gamma_R:~~c_{\pi_0}=\frac{e^2N_c}{48\pi^2f_\pi}2\sqrt{2}(Q_d^2-Q_u^2) ~,
\end{align}
\item[$\bullet$] From the full WZW interactions \(\mathcal{L}_{\rm WZW}^{\rm full}\) 
\begin{align}
ad\gamma_R d\gamma_R:~~c_{\rm wzw}=\frac{e^2N_c}{48\pi^2f}4(Q_u^2\alpha_u+Q_d^2\alpha_d)~,
\end{align}
\end{itemize}
with $Q_{u,d}$ being the charges of up and down quarks under the $U(1)'$ gauge group.

Similar to previous subsection, we define the common factor  
\begin{align} 
CF \equiv \left<\gamma\gamma|a\,d\gamma_Rd\gamma_R|a\right> ~,
\end{align}
in the amplitude to account for $a d\gamma_R d\gamma_R $. Combining the auxiliary rotation parameters in the three pieces together, one has 
\begin{align}
&\mathcal{M}(a\to \gamma_R \gamma_R)\text{(auxiliary)}=CF \times (c_{\rm ano} + c_{\rm WZW} + \theta_{a-\pi_0}' c_{\pi_0}) \\
&= CF \times e^2\left[\frac{-N_c}{48\pi^2f}2(Q_u^2\alpha_u+Q_d^2\alpha_d)+i\frac{f_\pi}{2\sqrt{2}f}\left((\alpha_u-\alpha_d)p_a^2-2\frac{m_u\alpha_u-m_d\alpha_d}{m_u+m_d}m_\pi^2 \right) \frac{i}{p_a^2-m_\pi^2} \times \frac{N_c}{48\pi^2f_\pi}2\sqrt{2}(Q_d^2-Q_u^2)\right] ~. \nonumber
\end{align}
There are several points to remark. Firstly, we have \(c_{\rm WZW} \neq 0\), and it takes up a similar form to \(c_{\rm ano}\). Combining these two terms leads to the coefficient \( -6 + 4 = -2\). Secondly, the chiral fundamental gauge boson has \(c_{\pi_0} \propto 2 \sqrt{2}\) instead of \(6 \sqrt{2}\) as in the \(a-\gamma -\gamma\) case, which can be told from the coupling of $\pi_0$ to the fundamental gauge fields in the WZW term $\Gamma_{\rm WZW}\supset id\pi_0(d\mathcal{A}_L\mathcal{A}_L+d\mathcal{A}_L\mathcal{A}_R+d\mathcal{A}_R\mathcal{A}_R)$. Thirdly, the amplitude contains auxiliary parameters not only in terms of the chiral phases \(\kappa_{u,d}\) but also the vector phases $\delta_{u,d}$, and the axion-pion mixing angle is also modified accordingly.  Using a similar breakup formula as Eq.~\eqref{eq:linearsplit}, along with the constraint \(\alpha_u + \alpha_d = 0\)~\footnote{This constraint can equivalently avoid regenerating the $aG\tilde{G}$ term, as by imposing \(\kappa_u + \kappa_d = 0\).} and the on-shell condition, we can verify that \(\mathcal{M}(a \to \gamma_R \gamma_R)\text{(auxiliary)} = 0\).

Similarly, we can obtain the final physical amplitude for $a d\gamma_R d\gamma_R$, $\mathcal{M}(a\to \gamma_R \gamma_R) = CF\times c_{\gamma_R \gamma_R}^{\rm eff}$, with the coefficient given by
\begin{align}
c_{\gamma_R \gamma_R}^{\rm eff} & = c_{\gamma_R\gamma_R}^0+(c_{\rm ano}+c_{\rm wzw})\frac{Q_u^2+Q_d^2}{2(Q_u^2 \alpha_{u,0}+Q_d^2 \alpha_{d,0})}-c_{\pi_0}\frac{f_\pi}{2\sqrt{2}f}\left(\frac{m_a^2}{m_\pi^2-m_a^2}\frac{\Tr[\tau_3(-\kRo)]}{2}-c_{gg}\frac{m_u-m_d}{m_u+m_d}\frac{m_\pi^2}{m_a^2-m_\pi^2}\right)\nonumber\\
&\quad +\frac{e^2N_c}{48\pi^2f}4(c_uQ_u^2+c_dQ_d^2)\nonumber\\
&=c_{\gamma_R\gamma_R}^0+e^2\left\{\frac{-N_cc_{gg}}{48\pi^2f}(Q_u^2+Q_d^2)+\frac{1}{2\sqrt{2}f}\left[\frac{m_a^2}{m_\pi^2-m_a^2}\left(\frac{c_u-c_d}{2}\right)+c_{gg}\frac{m_u-m_d}{m_u+m_d}\frac{m_\pi^2}{m_a^2-m_\pi^2}\right]\right.\nonumber\\
&\left.\quad \times \frac{N_c}{48\pi^2}2\sqrt{2}(Q_d^2-Q_u^2)\right\}+\frac{e^2N_c}{48\pi^2f}4(c_uQ_u^2+c_dQ_d^2) ~.
\end{align}

\subsubsection{$a$-$\omega$-$\gamma$ and $a$-$\rho$-$\gamma$ }
\label{app:a-omega-gamma}

In this subsection, we provide the first consistent calculations for the $a$-$\omega$-$\gamma$ and $a$-$\rho$-$\gamma$ scattering amplitudes. These amplitudes contain one fundamental gauge boson and one background/composite vector meson, and both couple to left and right-handed quarks equally. After the chiral basis transformation, we again trace the auxiliary rotation parameters in each of these contributions:
\begin{itemize}
\item[$\bullet$] From the anomaly contribution $ \delta \mathcal{L}_{\rm WZW}^{\rm ano}=-\delta[\mathcal{L}_{\rm WZW}+\mathcal{L}_c]$ that comes from the second term of Eq.~\eqref{eq:anomalous}:
\begin{align}
ad\omega d\gamma:~~c_{\rm ano}=\frac{-e g' N_c}{48\pi^2f}6(Q_u\kappa_u+Q_d\kappa_d) ~,
\end{align}
\item[$\bullet$] From the $\pi_0$ contribution 
\begin{align}
\pi_0d\omega d\gamma:~~c_{\pi_0}=\frac{e g'N_c}{48\pi^2f_\pi}6\sqrt{2}(Q_d-Q_u) ~,
\end{align}
\item[$\bullet$] From the full WZW interactions $\mathcal{L}_{\rm WZW}^{\rm full}$
\begin{align}
ad\omega d\gamma:~~c_{\rm wzw}=\frac{-e g' N_c}{48\pi^2f}6(Q_u\kappa_u+Q_d\kappa_d) ~.
\end{align}
\end{itemize}

We combine the auxiliary rotation parameters in the three pieces together to obtain
\begin{align}
&\mathcal{M}(a\to \omega\gamma)\text{(auxiliary)}= CF \times (c_{\rm ano} + \theta_{a-\pi_0}' c_{\pi_0} + c_{\rm wzw}) \\
&= CF \times e g'\left[\frac{-N_c}{48\pi^2f}12(Q_u\kappa_u+Q_d\kappa_d)+i\frac{f_\pi}{\sqrt{2}f}((\kappa_u-\kappa_d)p_a^2-2\frac{m_u\kappa_u-m_d\kappa_d}{m_u+m_d}m_\pi^2)\frac{i}{p_a^2-m_\pi^2} \times\frac{N_c}{48\pi^2f_\pi}6\sqrt{2}(Q_d-Q_u)\right] ~. \nonumber
\end{align}
Using a similar breakup formula as in Eq.~\eqref{eq:linearsplit}, along with the constraint \(\kappa_u + \kappa_d = 0\) and the on-shell condition, we can verify that \(\mathcal{M}(a \to \omega \gamma)\text{(auxiliary)} = 0\).

Similarly, we can obtain the final physical amplitudes for $a d\omega d\gamma$ and $a d\rho d\gamma$, $\mathcal{M}(a\to \omega \gamma) = CF\times c_{\omega \gamma}^{\rm eff}$ and $\mathcal{M}(a\to \rho \gamma) = CF\times c_{\rho \gamma}^{\rm eff}$.
The physical coefficient of $ad\omega d\gamma$ is given by 
\begin{align}
c_{\omega \gamma}^{\rm eff} & = \frac{(Q_u+Q_d)(c_{\rm wzw}+c_{\rm ano})}{2(Q_u\kappa_{u,0}+Q_d\kappa_{d,0})}-c_{\pi_0}\frac{f_\pi}{\sqrt{2}f}\left(\frac{m_a^2}{m_\pi^2-m_a^2}\frac{\Tr[\tau_3(\kLo-\kRo)]}{2}-c_{gg}\frac{m_u-m_d}{m_u+m_d}\frac{m_\pi^2}{m_a^2-m_\pi^2}\right)+\frac{eg'N_c}{48\pi^2f}(c_d+c_Q-2c_u)\nonumber\\
&=e g'\left\{\frac{-N_cc_{gg}}{48\pi^2f}6(Q_u+Q_d)+\frac{1}{\sqrt{2}f}\left[\frac{m_a^2}{m_\pi^2-m_a^2}\left(\frac{c_u-c_d}{2}\right)+c_{gg}\frac{m_u-m_d}{m_u+m_d}\frac{m_\pi^2}{m_a^2-m_\pi^2}\right]\times \frac{N_c}{48\pi^2}6\sqrt{2}(Q_d-Q_u)\right\} \nonumber \\
&\quad +\frac{eg'N_c}{48\pi^2f}(c_d+c_Q-2c_u) ~.
\end{align}
Substituting $N_c=3$, $Q_u=2/3$, $Q_d=-1/3$, we arrive at the final result, 
\begin{align}
c_{\omega \gamma}^{\rm eff} = e g'\left\{\frac{-c_{gg}}{8\pi^2f}-\frac{3}{8\pi^2f}\left[\frac{m_a^2}{m_\pi^2-m_a^2}\left(\frac{c_{u}-c_{d}}{2}\right)+c_{gg}\frac{m_u-m_d}{m_u+m_d}\frac{m_\pi^2}{m_a^2-m_\pi^2}\right]+\frac{1}{16\pi^2f}(c_d+c_Q-2c_u)\right\} ~,
\label{eq:a-omega-gamma}
\end{align}
which is the consistent physical coefficient independent of the auxiliary rotation parameters. 
Without providing a proof of the auxiliary rotation parameter cancellation, we simply write down the consistent physical coefficient for $a$-$\rho$-$\gamma$
\begin{align}
    c_{\rho\gamma}^{\rm eff}
&= eg\left\{ \frac{-3c_{gg}}{8\pi^2 f} - \frac{1}{8\pi^2f}\left[\frac{m_a^2}{m_\pi^2-m_a^2}\left(\frac{c_u-c_d}{2}\right) +c_{gg}\frac{m_u-m_d}{m_u+m_d}\frac{m_\pi^2}{m_a^2-m_\pi^2}\right] + \frac{1}{16\pi^2f}\left(3c_Q-2c_u-c_d\right) \right\} ~.
\label{eq:rhogamma}
\end{align}

\subsubsection{$a$-$\gamma$-$Z$}

The physical amplitude of $a$-$\gamma$-$Z$ has been calculated by using the quark-level effective Lagrangian at high energy in Ref.~\cite{Bauer:2021mvw}. It is intriguing to calculate the same amplitude using the full chiral axion Lagrangian at low energy.
The process contains two fundamental gauge bosons, one is vector-like and the other is chiral. In this case, besides the chiral auxiliary phase $\kappa_{u,d}$, the vector auxiliary phase $\delta_{u,d}$ dependence could also show up. The WZW and its counter-term have important contributions to cancel the auxiliary rotation phase dependence from the other parts. 

We directly use $Q_u=2/3$ and $Q_d=-1/3$ in the following calculations. Under the auxiliary chiral basis transformation, we trace the auxliary rotation parameters in the following three terms:
\begin{itemize}
\item[$\bullet$] From the WZW anomaly contribution $\delta \mathcal{L}_{\rm WZW}^{\rm ano}$ 
\begin{align}
ad\gamma dZ:~~c_{\rm ano}=\frac{N_c}{48\pi^2f}\frac{2e^2}{3c_ws_w}\left[3\delta_d+6\delta_u-3\kappa_d-6\kappa_u+4s_w^2(\kappa_d+4\kappa_u)\right] ~,
\end{align}
\item[$\bullet$] From the $\pi_0$ contribution 
\begin{align}
\pi_0d\gamma dZ:~~c_{\pi_0}=\frac{-e^2 N_c}{48\pi^2f_\pi s_wc_w}\sqrt{2}(c_w^2-3 s_w^2) ~,
\end{align}
\item[$\bullet$] From the full WZW interactions $\mathcal{L}_{\rm WZW}^{\rm full}$
\begin{align}
ad\gamma dZ:~~c_{\rm wzw}=\frac{-2e^2 N_c}{48\pi^2f s_w c_w}(\delta_d+2\delta_u) ~.
\end{align}
\end{itemize}

We combine the auxiliary rotation parameters in the three pieces to obtain
\begin{align}
&\mathcal{M}(a\to Z^*\gamma)\text{(auxiliary)}=CF \times (c_{\rm ano} + \theta_{a-\pi_0 }'\,c_{\pi_0} + c_{\rm wzw}) \\
&= CF \times \left[c_{\rm wzw}+c_{\rm ano}+i\frac{f_\pi}{\sqrt{2}f}\left((\kappa_u-\kappa_d)p_a^2-2\frac{m_u\kappa_u-m_d\kappa_d}{m_u+m_d}m_\pi^2 \right) \frac{i}{p_a^2-m_\pi^2} \times c_{\pi_0}\right].
\end{align}

It is clear that the auxiliary vector phases \(\delta_{u, d}\) are only contained in \(c_{\rm ano}\) and \(c_{\rm WZW}\), and they cancel each other out. Similar to the previous calculations, by imposing the constraint \(\kappa_u + \kappa_d = 0\) and the on-shell condition, we can explicitly write down the auxiliary rotation parameters in the amplitude as
\begin{align}
\mathcal{M}(a\to Z^*\gamma)\text{(auxiliary)} & \propto \frac{e^2N_c}{48\pi^2fs_wc_w}\left\{(-2+2)(\delta_d+2\delta_u)
+\frac{2}{3}\left[-3\kappa_d-6\kappa_u+4s_w^2(\kappa_d+4\kappa_u)\right]+(c_w^2-3s_w^2)(\kappa_u-\kappa_d)\right\} \nonumber\\
&=\frac{e^2N_c}{48\pi^2fs_wc_w}\left\{(\kappa_u-\kappa_d)(-1+4s_w^2+c_w^2-3s_w^2)\right\}=0 ~,
\end{align}
which shows that they indeed cancel out.

Finally, we arrive at the physical result for the coefficient of the $a$-$\gamma$-$Z$ scattering, which is given by
\begin{align}
c_{\gamma Z}^{\rm eff}=&c_{\gamma Z}^0+\frac{N_c c_{gg}}{48\pi^2f}\frac{e^2}{s_wc_w}(-9+20s_w^2)-c_{\pi_0}\frac{f_\pi}{\sqrt{2}f}\left(\frac{m_a^2}{m_\pi^2-m_a^2} \frac{c_d-c_u}{2} -c_{gg}\frac{m_u-m_d}{m_u+m_d}\frac{m_\pi^2}{m_a^2-m_\pi^2}\right)\nonumber\\
&-\frac{N_c}{48\pi^2f}\frac{2e^2}{s_{2w}}(c_d+2c_u+3c_Q) ~,
\label{eq:a-Z-gamma}
\end{align}
where $c_{\gamma Z}^0$ comes from the contribution at the UV scale.

The \(a\)-\(\gamma\)-\(Z\) interaction may have several applications. In a neutrino dense environment, the \(Z\) boson can be interpreted as the background lepton current. This interaction can induce a Primakov-like conversion \(a \nu \leftrightarrow \gamma \nu\), which could be of interest for supernova physics. In addition, for a light axion (\(m_a \ll 2\,\text{GeV}\)), it may induce an exotic decay through an off-shell \(Z\), \(a \to Z^* \gamma \to \bar{\nu} \nu \gamma\). The partial width can be calculated and shown to be
\begin{align}
\Gamma_{a \to \nu \bar{\nu} \gamma} = \frac{3 \alpha\,m_a^7}{5 \cdot 2^7 \pi^2 \sin(2\theta_{\rm w})^2 m_Z^4} |c_{\gamma Z}^{\rm eff}|^2 ~.
\end{align}

\subsubsection{$a$-$Z$-$\omega$}
The $a$-$Z$-$\omega$ amplitude contains one fundamental chiral gauge boson and one vector-like background gauge boson. Similar to the $a$-$Z$-$\gamma$ calculation,  we choose $Q_u=2/3$ and $Q_d=-1/3$. Under the auxiliary chiral basis transformation, we trace the auxiliary rotation parameters in the following three terms:
\begin{itemize}
\item[$\bullet$] From the WZW anomaly contribution $\delta \mathcal{L}_{\rm WZW}^{\rm ano}$ 
\begin{align}
adZ d\omega:~~c_{\rm ano}=\frac{eg'N_c}{48\pi^2fs_{2 {\rm w}}}\left[-3\delta_d+3\delta_u+3\kappa_d-3\kappa_u+s_{\rm w}^2(-4\kappa_d+8\kappa_u)\right] ~,
\end{align}
\item[$\bullet$] From the $\pi_0$ contribution 
\begin{align}
\pi_0dZ d\omega:~~c_{\pi_0}=\frac{eg' N_c}{48\pi^2f_\pi s_{\rm w}c_{\rm w}}3\sqrt{2}(-1+2s_w^2) ~,
\end{align}
\item[$\bullet$] From the full WZW interactions $\mathcal{L}_{\rm WZW}^{\rm full}$
\begin{align}
adZ d\omega:~~c_{\rm wzw}=\frac{eg'N_c}{48\pi^2fs_{2{\rm w}}}\left[3\delta_d-3\delta_u+3\kappa_d-3\kappa_u+s_{\rm w}^2(-4\kappa_d+8\kappa_u)\right] ~.
\end{align}
\end{itemize}
We combine the auxiliary rotation parameters in the three pieces to obtain
\begin{align}
&\mathcal{M}(a\to Z^*\omega)\text{(auxiliary)}= CF \times (c_{\rm ano} + \theta_{a-\pi_0 }'c_{\pi_0} + c_{\rm wzw}) \\
&= CF \times \left[c_{\rm wzw}+c_{\rm ano}+i\frac{f_\pi}{\sqrt{2}f}\left((\kappa_u-\kappa_d)p_a^2-2\frac{m_u\kappa_u-m_d\kappa_d}{m_u+m_d}m_\pi^2\right)\frac{i}{p_a^2-m_\pi^2}\times c_{\pi_0}\right] ~.
\end{align}
Imposing the constraint $\kappa_u+\kappa_d=0$ and the on-shell condition, it is straightforward to show that the amplitude with auxiliary rotation parameters vanishes or $\mathcal{M}(a\to Z^*\omega)\text{(auxiliary)}=0$.

For reference and convenience, we provide the consistent physical result for the coefficient of $a$-$Z$-$\omega$ as
\begin{align}
c_{Z\omega}^{\rm eff}=&\frac{N_c\,c_{gg}}{48\pi^2f}\frac{2\,s_{\rm w}\,e\,g'}{c_{\rm w}}-c_{\pi_0}\frac{f_\pi}{\sqrt{2}f}\left(\frac{m_a^2}{m_\pi^2-m_a^2}\frac{c_d-c_u}{2}-c_{gg}\frac{m_u-m_d}{m_u+m_d}\frac{m_\pi^2}{m_a^2-m_\pi^2}\right)\nonumber\\
&+\frac{N_c}{48\pi^2f}\frac{e\,g'}{s_{2{\rm w}}}\left(3c_d-3c_u-2s_{\rm w}^2(c_d+c_Q-2c_u)\right) ~.
\label{eq:a-Z-omega}
\end{align}

\subsection{The connection to the vector meson dominance}
\label{app:vector-meson-dominance}

The principle of vector meson dominance (VMD) was proposed by Sakurai in a seminal paper~\cite{Sakurai:1960ju}. It essentially states that the interactions between hadrons and photons are predominantly mediated by vector mesons, such as $\rho_0$ and $\omega$, through their mixing with photons. By treating vector mesons as background fields, Ref.~\cite{Kaymakcalan:1983qq} derived the $\pi_0\to\gamma\gamma$ result using the VMD formalism and obtained the same result from $U(1)_{\rm EM}$ of WZW interactions. In the Hidden Local Symmetry framework~\cite{Bando:1987br}, the VMD matching for the $\pi_0\to\gamma\gamma$ process is achieved. VMD can also be extended to off-shell photons, leading to the electromagnetic form factors $\mathcal{F}(q^2)$ of the hadron-photon effective vertex. These electromagnetic form factors characterize the spatial distribution of charge for an extended object under high momentum transfer of photons. The form factors developed in this way agree well with experimental data~\cite{Schildknecht:2005xr,Lomon:2016eyp,Czerwinski:2012ry,Achasov:2013btb,SND:2023gan}. 
Therefore, it is intriguing to examine whether our full axion-interaction Lagrangian is compatible with the VMD principle. We will compare the results from VMD and $U(1)_{\rm EM}$ gauging on the physical amplitudes for $a$-$\gamma$-$\gamma$ and $a$-$\omega$-$\gamma$ as two concrete examples.

\begin{figure}[th!]
\centering
\includegraphics[width= 0.6 \linewidth]{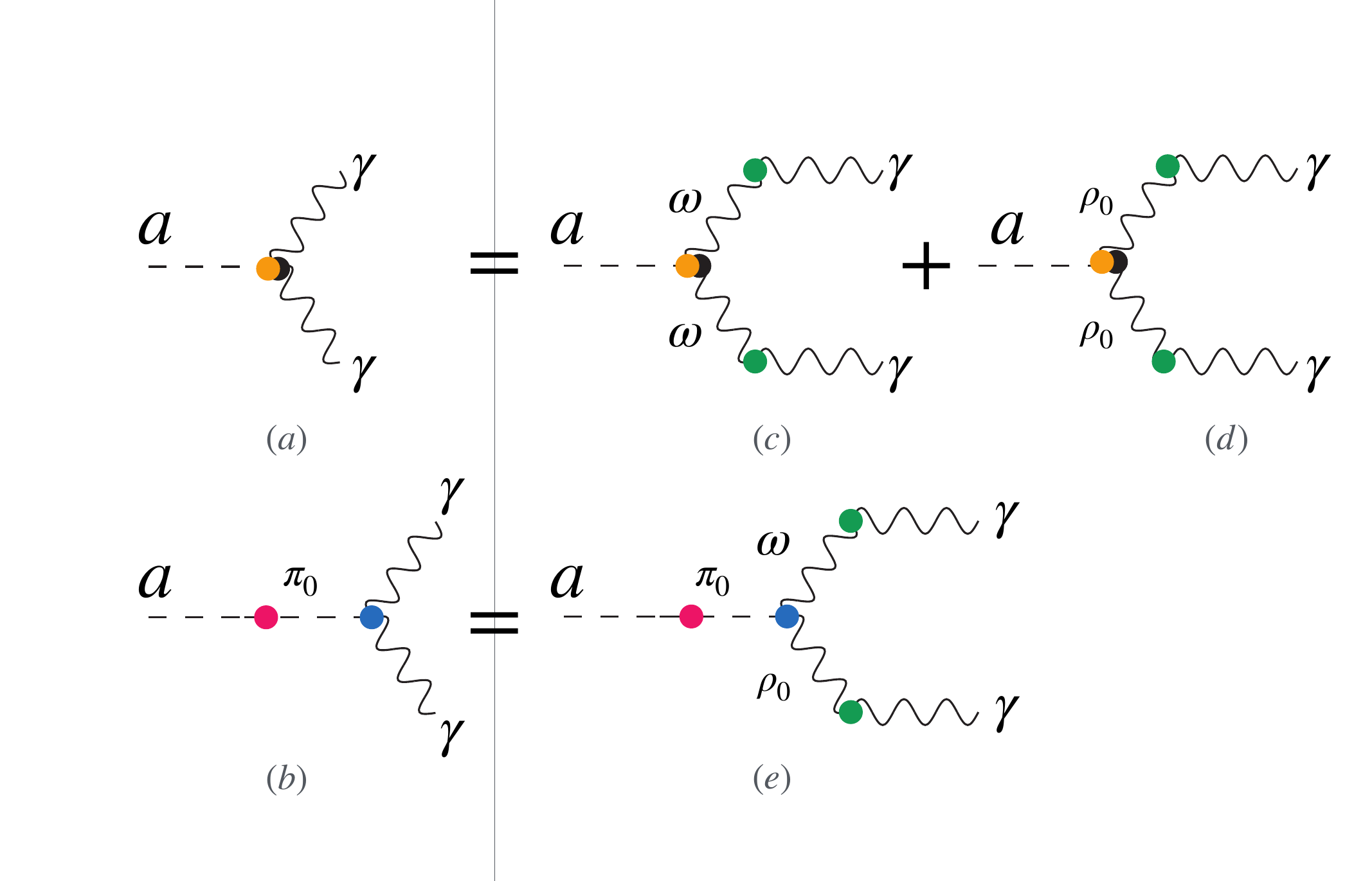}
\caption{The diagrams for the \(a\)-\(\gamma\)-\(\gamma\) interaction using the chiral axion Lagrangian are shown on the left, while the diagrams using the Vector Meson Dominance model are on the right. The red dot \textcolor{red}{$\bullet$} indicates the \(a\)-\(\pi_0\) mixing, the green dot {\color{ForestGreen} $\bullet$} indicates the $\gamma$ and $\omega/\rho$ mixing, the black dot \textcolor{black}{$\bullet$} indicates the interactions from \(\mathcal{L}_{\rm WZW}^{\rm full}\), the orange dot {\color{orange} $\bullet$}  indicates the interactions from the anomalous contribution \(\mathcal{L}_{\rm \chi PT}^{\rm ano}\), and the blue dot {\color{NavyBlue} $\bullet$} indicates the \(\pi_0\) interactions in the Standard Model. The diagrams with exchanged external photon legs are omitted. Diagram (a) equals diagram (c) plus (d), while diagram (b) equals diagram (e) trivially because VMD for \(\pi_0\) has already been proven in the SM. There is no $a$-$\rho$-$\omega$ diagram included because we consider a special $\mathcal{L}_{\rm eff, 0}$.}
\label{fig:VMD_agammagamma} 
\end{figure}

In the context of VMD, the photon solely mixes with vector mesons, and hadrons couple to photons through the mediation of virtual vector mesons. Deriving the physical effective vertex of $a\gamma\gamma$ using the VMD formalism requires considering the axion effective coupling with vector mesons, where the vector mesons are attached to an on-shell photon, as illustrated by the red dot in Fig.~\ref{fig:VMD_agammagamma}.
The photon-vector meson mixing terms are given by~\cite{Kaymakcalan:1983qq,Harada:2003jx}
\begin{align}
    \mathcal{L}_{\rm VMD}=\frac{e}{g'}\frac{m_\omega^2}{3}(A_\gamma)_\mu\omega^\mu+\frac{e}{g}m_\rho^2(A_\gamma)_\mu\rho_0^\mu+ \cdots ~,
\end{align}
where we have omitted terms related to other meson mixings.
Note that due to a convention difference, our notation has a factor of $\sqrt{2}$ difference compared to Ref.~\cite{Kaymakcalan:1983qq}, so in our work, we have $g' \sim g = f_{\rho\pi\pi} \approx \sqrt{12\pi}$.

The physical effective vertex of $a$-$\gamma$-$\gamma$ with on-shell photons is given by Eq.~\eqref{eq:agammagamma}. It comprises two contributions, which are shown on the left-hand side of Fig.~\ref{fig:VMD_agammagamma}. One part originates from the axion anomalous interactions (the upper left diagram), while the other part results from the axion mixing with $\pi_0$ (the lower left diagram). 

For the VMD formalism, we first calculate the axion-pion mixing part, which is the lower right term of Fig.~\ref{fig:VMD_agammagamma}. This one should trivially match the lower left term, because it is already proven that $\pi_0 \to \gamma\gamma$ matches to the VMD~\cite{Kaymakcalan:1983qq, Bando:1987br}. Nevertheless, we provide our demonstration as well for the calculation of the amplitude for the lower right diagram,
\begin{align}
i \frac{f_\pi}{\sqrt{2}f}\left[-\frac{m_u-m_d}{m_u+m_d}c_{gg}m_\pi^2\right]\frac{i}{p_a^2-m_\pi^2}i\left(\frac{-3\sqrt{2}gg'}{8\pi^2f_\pi}\right)\cdot \frac{ie}{g}m_\rho^2\left[D(m_\rho^2,p_1^2)D(m_\omega^2,p_2^2)+D(m_\rho^2,p_2^2)D(m_\omega^2,p_1^2)\right]\frac{ie}{3g'}m_\omega^2 ~,
 \label{eq:agamgam_part1}
\end{align}
with $D(m^2,p^2)\equiv i/(p^2-m^2)$ denoting the leading-order vector meson propagator. $p_1$ and $p_2$ are the momenta of the external photons. By taking the photons on-shell, $p_1^2 = p_2^2 = 0$, one can see that it matches the axion diphoton amplitude from the $\pi_0$ mixing in Eq.~\eqref{eq:agammagamma} after taking the $adAdA$ vertex symmetry factor 2 into account.

Besides the axion-pion mixing part, we also check the parts related to the axion couplings to the vector mesons, which are the $a$-$\omega$-$\omega$ and $a$-$\rho$-$\rho$ diagrams in the upper right of Fig.~\ref{fig:VMD_agammagamma}. Their amplitudes are given by
\begin{align}
    \left[ic_{\rho\rho}\cdot 2\cdot D(m_\rho^2,p_1^2)D(m_\rho^2,p_2^2)\left(\frac{ie}{g}m_\rho^2\right)^2\right]+\left[ic_{\omega\omega}\cdot 2\cdot D(m_\omega^2,p_1^2)D(m_\omega^2,p_2^2)\left(\frac{ie}{3g'}m_\omega^2\right)^2\right].
    \label{eq:agamgam_part2}
\end{align}
Here, the  $a$-$\omega$-$\omega$ and $a$-$\rho$-$\rho$ coefficients are given by
\begin{align}
    c_{\rho\rho}=\left[-\frac{2g^2c_{gg}}{16\pi^2f}(\text{full WZW})-\frac{g^2c_{gg}}{16\pi^2f}(\text{ano})\right]=-\frac{3g^2c_{gg}}{16\pi^2f} ~, \\
      c_{\omega\omega}=\left[-\frac{2g'^2c_{gg}}{16\pi^2f}(\text{full WZW})-\frac{g'^2c_{gg}}{16\pi^2f}(\text{ano})\right]=-\frac{3g'^2c_{gg}}{16\pi^2f} ~,
\end{align}
from the full WZW interactions plus the anomalous contribution when rotating out the $aG\tilde{G}$ term. The symmetry factor has been taken into account. There is no $a$-$\rho$-$\omega$ contribution due to the isospin symmetry, because we consider a special case in which the axion only interacts with the gluons in $\mathcal{L}_{\rm eff, 0}$.
After putting the two photon on-shell, we find that it matches the $a$-$\gamma$-$\gamma$ amplitude from the full WZW interactions plus the anomalous contribution when rotating out the $aG\tilde{G}$ term, which is the upper left diagram in Fig.~\ref{fig:VMD_agammagamma}.

\begin{figure}[thb!]
\centering
\includegraphics[width= 0.6 \linewidth]{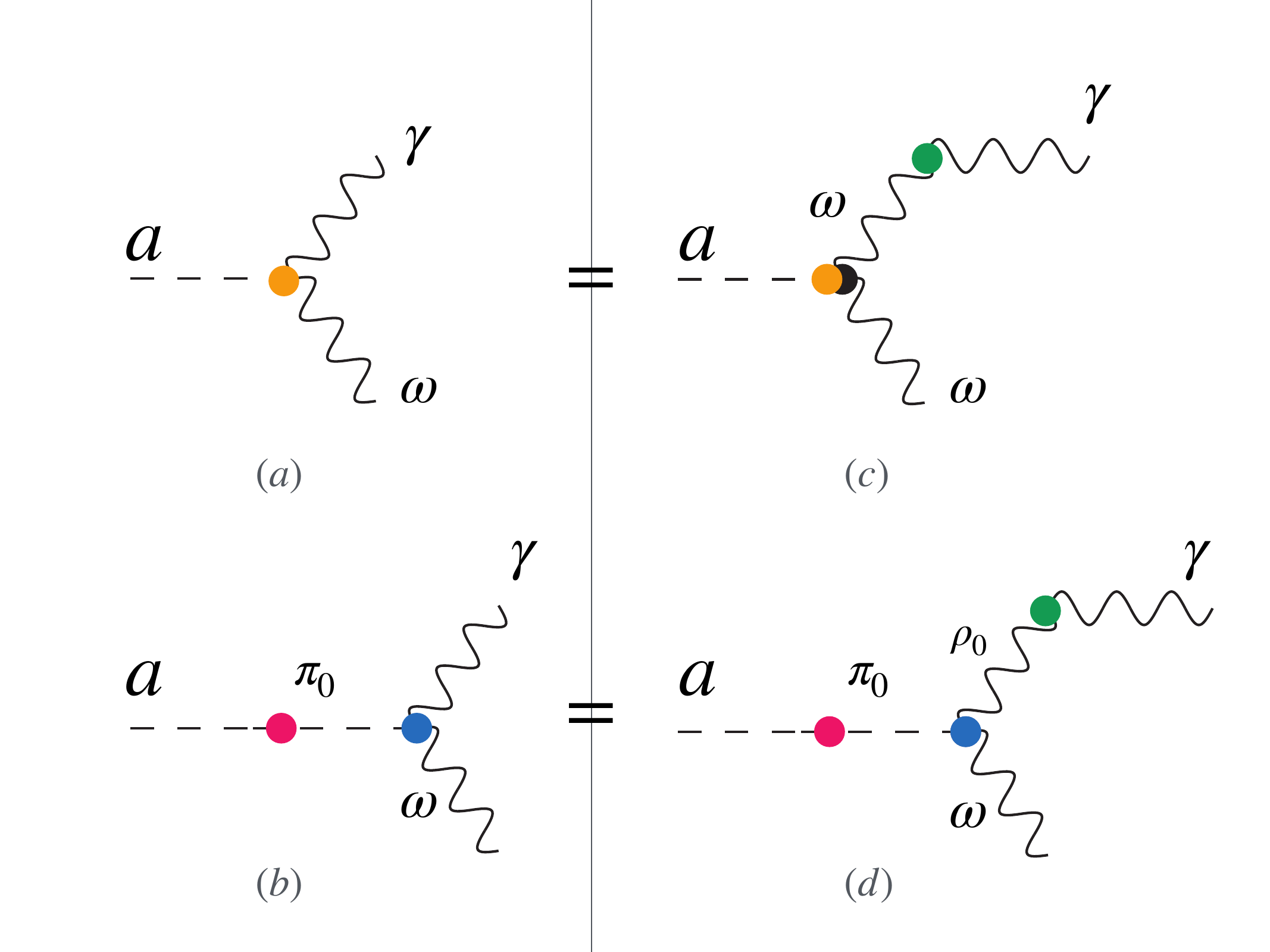}
\caption{The diagrams for the \(a\)-\(\gamma\)-\(\omega\) interaction using the chiral axion Lagrangian are shown on the left, while the diagrams using the Vector Meson Dominance model are on the right. The notations are similar to those in Fig.~\ref{fig:VMD_agammagamma}.}
\label{fig:VMD_aomegagamma} 
\end{figure} 

After calculating the VMD for the \(a\)-\(\gamma\)-\(\gamma\) interaction, we turn to another amplitude, \(a\)-\(\gamma\)-\(\omega\). The procedure is similar, and the diagrams are provided in Fig.~\ref{fig:VMD_aomegagamma}. The difference is that the upper right panel contains only one diagram from \(a\)-\(\omega\)-\(\omega\). In addition, we only need to include the photon-vector meson mixing once in each diagram.

The axion-pion mixing contribution from the VMD is given by
\begin{align}
i \frac{f_\pi}{\sqrt{2}f}\left[-\frac{m_u-m_d}{m_u+m_d}c_{gg}m_\pi^2\right]\frac{i}{p_a^2-m_\pi^2}i\left(\frac{-3\sqrt{2}gg'}{8\pi^2f_\pi}\right)D(m_\rho^2,p_2^2)\frac{ie}{g}m_\rho^2 \,,
 \label{eq:aomegagam_part1}
\end{align}
while the contribution from $a$-$\omega$-$\omega$ is given by
\begin{align}
ic_{\omega\omega}\cdot 2\cdot D(m_\omega^2,p_2^2)\left(\frac{ie}{3g'}m_\omega^2\right) \,,
    \label{eq:aomegagam_part2}
\end{align}
with a symmetry factor of 2.  
Again, by setting \(p_2^2 = 0\) and \(p_a^2 = m_a^2\), and substituting the explicit expression for \(c_{\omega \omega}\), the VMD results match with the direct calculations of the left-hand-side diagrams.

After the above calculations, we comment on the relationship between VMD and our chiral axion Lagrangian. In the above calculations, we see that the VMD model leads to the same results as the direct calculation using the chiral axion Lagrangian for the axion-gluon coupling. However, we have also checked that if the axion couples to the quark bilinear in \(\mathcal{L}_{\rm eff, 0}\), the direct calculation and the VMD calculation do not match.

The mismatch of quark bilinear coupling results should not be surprising. The reason is as follows: In the VMD model, all photon couplings arise from the vector meson mixing. Thus, the photon is treated equally as a vector meson. However, in the chiral axion Lagrangian, vector bosons are classified into two categories: fundamental gauge bosons and background gauge bosons (or vector mesons). We must satisfy the gauge invariance for the fundamental gauge symmetries but not for the gauge symmetries of the background gauge bosons. Following this principle, additional counter-terms must be added to the WZW interactions. Therefore, the fundamental gauge bosons, such as photon, are treated differently from the background vector bosons, such as the \(\omega\) or \(\rho\) mesons. 
Additionally, it is important to note that the pion meson is a composite particle made of quarks, while the axion is a fundamental pseudoscalar.  Therefore, it should not be surprising if VMD matching fails for the quark couplings of the axion, even though the hidden local symmetry model that effectively describes interactions between composite pseudoscalar and vector mesons successfully predicts VMD for the pion meson.

However, the match for the axion-gluon coupling \(c_{gg}\) is successful, which is non-trivial and very surprising. For example, in the \(a\)-\(\gamma\)-\(\gamma\) calculation in Fig.~\ref{fig:VMD_agammagamma}, the direct chiral axion Lagrangian calculation only involves the anomalous contribution because no WZW contribution to \(a\)-\(\gamma\)-\(\gamma\) exists. However, in the VMD calculation, there are WZW contributions involved for \(a\)-\(\omega\)-\(\omega\) and \(a\)-\(\rho\)-\(\rho\). This match spans very different types of interactions. We still do not understand why this happens or if there is a deeper reason. We postpone this investigation to future work. 

\section{The collider searches via new axion WZW interactions}

With the new WZW interactions for axions, we will explore new phenomenology in collider searches. We will focus on low-energy electron-positron colliders, specifically the Beijing Spectrometer (BESIII) \cite{BESIII:2022dxl} and the proposed Super-Tau-Charm Facility (STCF) \cite{Achasov:2023gey} with low center-of-mass energies, because the chiral Lagrangian breaks down at high energy. There are many couplings for axions in the effective Lagrangian at the quark level. We choose to focus on the axion-gluon coupling and neglect the other quark couplings, which provides a simple model to start with.

\subsection{The decay of axion}
\label{app:axion-decay-width}

\begin{figure}{}
\centering
\includegraphics[width= 0.49 \linewidth]{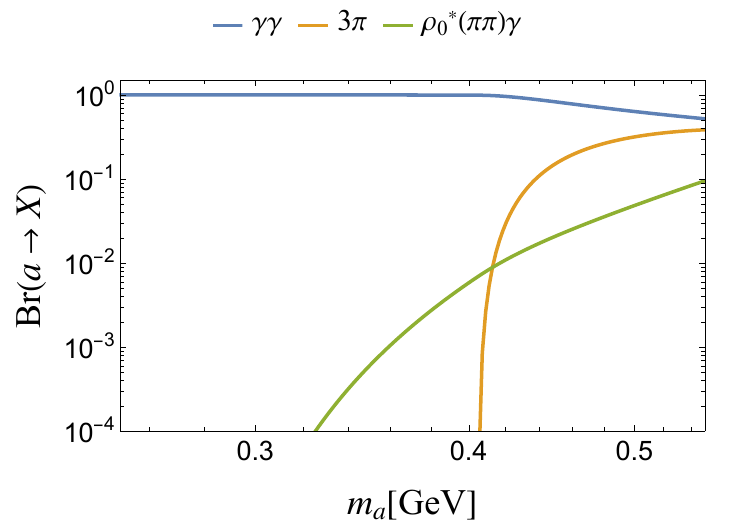}
    \includegraphics[width= 0.49 \linewidth]{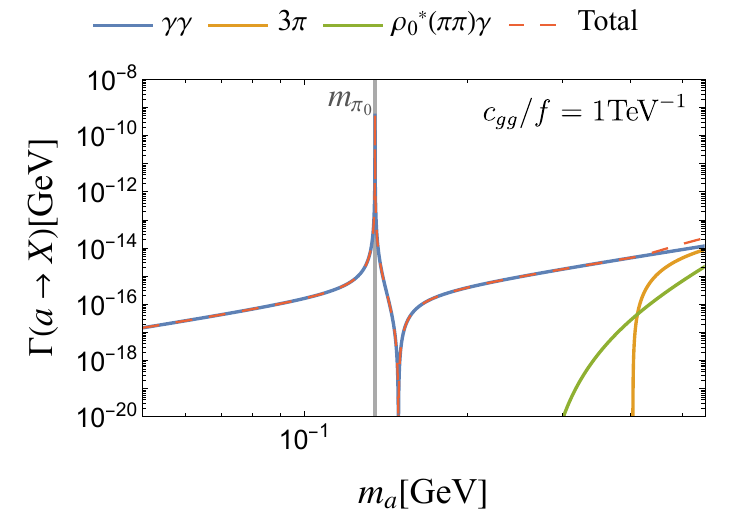}
\caption{\textit{Left panel}: The axion decay branching ratios for $m_a < m_\eta\sim 550{\rm MeV}$, assuming the axion only couples to gluons in the effective quark-level Lagrangian.
\textit{Right panel}: The partial decay width of various decay modes. }
\label{fig:axionbr} 
\end{figure} 

We will study the new production channels for axions and explore its implications. Before looking into the production channels, we will first calculate the axion decay channels and various partial widths. For a low energy below the QCD scale, the axion prodominalty decays into scalar mesons, vector mesons, and photons. Besides decaying to two photons, an axion could also have numerous hadronic decay modes~\cite{Aloni:2018vki}. To avoid the complexity of axion hadronic decay modes and the mixing effect with the $\eta(\eta')$ mesons, we restrict our analysis to \(m_a \lesssim 500 \, \text{MeV}\) for simple analytic calculations.

For axion decays into pions, due to CP invariance and angular momentum conservation, the dominant decay channels are \(a \to \pi^+\pi^-\pi^0\) and \(a \to 3 \pi^0\). The three-body decay mode \(a \to \pi^0 \gamma \gamma\) is further suppressed by additional powers of the fine structure constant \(\alpha\). Therefore, it is much smaller compared to the \(3 \pi\) channel and can be neglected~\cite{Bauer:2017ris, Bauer:2021mvw}. 

However, the decay channel \(a \to \pi^+\pi^-\gamma\) could have significant contributions from \(a \to \gamma \rho^* (\pi^+\pi^-)\), with the \(\rho\) being off-shell. 
We do not consider the four-body decay final states due to strong phase-space suppression. In summary, for \(m_a < 500 \, \text{MeV}\), the major decay modes are \(\gamma\gamma\), \(3\pi\), and \(\pi^+\pi^-\gamma\).
The partial widths of the three channels are
\begin{align}
   \Gamma_{a \to \gamma \gamma} & = \frac{m_a^3}{16\pi}\frac{|c_{\gamma\gamma}^{\rm eff}|^2}{f^2} ~, \\
    \Gamma_{a \to 3 \pi} & = \frac{\pi}{24}\frac{m_am_\pi^4}{(16\pi^2)^2f^2f_\pi^2}(\Delta_{ud})^2\left[g_{00}\left(\frac{m_\pi^2}{m_a^2}\right)+g_{+-}\left(\frac{m_\pi^2}{m_a^2}\right)\right] ~, \\
    \Gamma_{a\to\pi^+\pi^-\gamma} & = \frac{g^2|c_{\rho\gamma}^{\rm eff}|^2}{3(8\pi)^3}m_a^3\int_{4m_\pi^2}^{m_a^2}dm_{12}^2~ m_{12}^2\left(1-\frac{4m_\pi^2}{m_{12}^2}\right)^{\frac{3}{2}}\left|{\rm BW}(m_{12}^2)\right|^2\left(1-\frac{m_{12}^2}{m_a^2}\right)^3 ~,
\end{align}
with coefficients $c_{\gamma\gamma}^{\rm eff}$ and $c_{\rho\gamma}^{\rm eff}$ given in Eqs.~\eqref{eq:agammagamma} and \eqref{eq:rhogamma}, respectively. The parameter $\Delta_{ud}= 2(c_u-c_d)+2 c_{gg}\frac{m_d-m_u}{m_d+m_u}$, with $m_u/m_d \simeq 0.46$, and the function BW represents the Breit-Wigner function for the $\rho_0$ meson.
The phase space functions $g_{00}(r)$ and $g_{+-}(r)$ are valid within $0 \leq r \leq 1/9$ and are provided in Refs.~\cite{Bauer:2017ris,Bauer:2021mvw} as follows:
\begin{align}
g_{00}(r)&=\frac{2}{(1-r)^2}\int_{4r}^{(1-\sqrt{r})^2}dz\sqrt{1-\frac{4r}{z}}\lambda^{1/2}(z,r) ~, \\
g_{+-}(r)&=\frac{12}{(1-r)^2}\int_{4r}^{(1-\sqrt{r})^2}dz\sqrt{1-\frac{4r}{z}}(z-r)^2\lambda^{1/2}(z,r) ~, 
\end{align}
with the intermediate function $\lambda(z,r) \equiv (1-z-r)^2-4 z r$. 

We show the branching ratios and the partial decay widths of the three dominant axion decay channels in Fig.~\ref{fig:axionbr}. In the right panel, the peak at \(m_a \sim m_{\pi^0}\) indicates the maximum axion-pion mixing, which greatly enhances the effective \(c_{\gamma\gamma}^{\rm eff}\) decay coefficient. However, in this region, the perturbativity of the axion-pion interaction breaks down, and a more proper approach is needed to diagonalize the axion-pion \(2 \times 2\) matrix.
For \(m_a < 550\) MeV, it is clear that the diphoton decay channel dominates. For a higher \(m_a\), the full axion chiral Lagrangian should be capable to calculate the hadronic decay partial widths. Given that there are too many hadronic final states, we leave this analysics to future study.

\subsection{The phenomenology at a low energy $e^+e^-$ collider}
\label{app:pheno-ee-collider}

Given the new interactions among axion, (axial-)vector mesons, and photons, there are many potential applications at low-energy colliders such as the Beijing Spectrometer (BESIII)~\cite{BESIII:2022dxl} and the proposed Super-Tau-Charm Facility (STCF)~\cite{Achasov:2023gey}, which operate at center-of-mass energies around 4 GeV. In the past few years, BESIII has accumulated data with an integrated luminosity of $4785.35~{\rm pb}^{-1}$ at center-of-mass energies ranging from $ 4090$ to $4600$ MeV~\cite{BESIII:2022dxl}, while the STCF has planned a scan run with center-of-mass energies from $3.097$ to $4.630$ GeV and an integrated luminosity of $10~\text{ab}^{-1}$~\cite{Achasov:2023gey}. One could also consider even lower energy colliders like SND~\cite{Aulchenko:2009zz}, though its integrated luminosity is much smaller than that of BESIII and STCF. Additionally, our analysis is restricted to a hadronphilic QCD-like axion, which only has a gluon coupling \(c_{gg}\) within \(\mathcal{L}_{\rm eff, 0}\).

\subsubsection{The \(e^+e^- \to \omega + a\) cross-section}
\label{app:omega-a-cross-section}

We investigate the associated production of a vector meson and an axion via the process \(e^+e^- \to \omega + a\), a scenario  not previously studied. Similarly, one could also study the process of \(e^+e^- \to \rho_0 + a\).
Depending on the axion's lifetime and mass, several detection scenarios arise. If the axion is light and flies out of the detector before decaying, it leads to the semi-visible signature, \(\omega + \text{invisible}\). If the light axion decays within the detector after traveling a certain distance, the signature contains a displaced diphoton vertex, \(\omega + \text{displaced vertex}\). Lastly, if the axion is heavy and decays promptly, it can decay into the three channels shown in Fig.~\ref{fig:axionbr}, with the dominant channel being \(a \to \gamma \gamma\).

Within the framework of the chiral axion Lagrangian, we can calculate the cross-section for \(e^+e^- \to \omega + a\) using the vertex coefficient \(c_{\omega \gamma}^{\rm eff}\). This process is mediated by an s-channel photon, which is off-shell with \(q^2 = s\), the square of the center-of-mass energy of the collider. To account for an off-shell photon, we incorporate the Vector Meson Dominance (VMD) model to handle the form factor suppression.

Given that we have established VMD's applicability to the \(a\)-\(\gamma\)-\(\omega\) vertex for the \(c_{gg}\) coupling, the VMD procedure can provide the necessary form factors. Thus, the s-channel photons can be substituted to vector mesons through mixing, to couple to the axion. For the mediation of \(\omega\) mesons, which have a decay width that adheres to the narrow width approximation, we use the Breit-Wigner function. For the \(\rho_0\) meson, which has a larger decay width and includes various excited states \(\rho_i\), we adopt the form factors from Ref.~\cite{SND:2023gan}, derived by fitting the VMD formula to the Born cross-section obtained from experimental data.

As a result, we obtain the effective coefficients \(c_{\omega \gamma}^{\rm eff}(q^2)\) for the off-shell \(\gamma^*(q^2 > 0)\), on-shell axion \(a\), and \(\omega\) meson as
\begin{align}
    c_{\omega \gamma}^{\rm eff}(q^2)=-\frac{eg'c_{gg}}{8\pi^2f}\frac{m_\omega^2}{m_\omega^2-q^2-i\sqrt{q^2}\Gamma_\omega} - \frac{3eg'c_{gg}}{8\pi^2f}\frac{m_u-m_d}{m_u+m_d}\frac{m_\pi^2}{m_a^2-m_\pi^2}\sum_{i=0}^{3}\frac{A_i \, M_i^2\,e^{i\phi_i}}{M_i^2-q^2-i\sqrt{q^2}\Gamma_i(\sqrt{q^2})} ~,
\end{align}
where the first term arises from \(\omega\) mediation, while the second term is due to the mediation through \(\rho_0\) and its excited states. The \(\rho\)-meson-related Breit-Wigner function parameters \(M_i\), \(A_i\), \(\phi_i\), and \(\Gamma_i\) can be found in Ref.~\cite{SND:2023gan}. 

Thus, the differential cross-section for \(e^+e^- \to \omega + a\) is given by
\begin{align}
\frac{d\sigma(e^+e^-\to \omega~a)}{d\cos\theta}=\frac{\alpha|c_{\omega\gamma}^{\rm eff}(q^2)|^2 \left[m_a^4+(m_\omega^2-s)^2-2m_a^2(m_\omega^2+s)\right]}{64 f^2s^2}(1+\cos\theta^2) ~,
\end{align}
with $\theta$ denoting the angle between the axion momentum and the beam direction. 
In principle, the \(e^+e^- \to \gamma^* + a\) process could have a possible contribution from the \(a\)-\(\gamma\)-\(\gamma\) vertex, where one substitutes the outer leg of the photon with an \(\omega\) meson through \(\omega\)-\(\gamma\) mixing in the VMD framework~\cite{Davier:2006fu, Lu:2024cjp}. However, this additional contribution is suppressed compared to the \(a\)-\(\omega\)-\(\gamma\) vertex by the coupling difference \((e/3g') \ll 1\). Therefore, we do not take this additional contribution into consideration.

\subsubsection{$\omega + {\rm invisible}$}
\label{app:omega+invisible}

For an axion with a long lifetime, it can decay outside the detector. Therefore, its signature at the collider is mono-\(\omega\) plus missing energy. In this case, the background events primarily arise from two categories: one involves \(e^+e^- \to \omega \gamma\) with a photon mistagged as an invisible object. For photons escaping detection due to the limited coverage of the detector, one can apply cuts to avoid such regions. Thus, photon mistagging to missing energy constitutes one category of the background. The other is the irreducible SM background \(e^+e^- \to \omega Z^*(\bar{\nu}\nu)\) induced by the $Z$-$\gamma$-$\omega$ interaction in the WZW term.

Regarding the photon mistagged background, the \(e^+e^- \to \omega \gamma\) cross-section can be calculated via VMD~\cite{Davier:2006fu,Lu:2024cjp}:
\begin{align}
\frac{d\sigma_{V\gamma}}{d\cos\theta}&=\left(\frac{e}{3g'}\right)^2\frac{d\sigma_{\gamma_1^*\gamma_2^*}}{d\cos\theta}(m_\omega^2,0)~,\nonumber\\
\frac{d\sigma_{\gamma_1^*\gamma_2^*}}{d\cos\theta}(m_{V_1}^2,m_{V_2}^2)&=\frac{\pi\alpha^2}{s}\frac{2|\vec{p}|}{\sqrt{s}}\frac{2(m_{V_1}^2+m_{V_2}^2)s\,u\,t+(t^2+u^2)(u\,t-m_{V_1}^2m_{V_2}^2)}{u^2\,t^2} ~,
\end{align}
with \(\vec{p}\) being the meson momentum in the center-of-mass frame, and \(s\), \(t\), and \(u\) being the Mandelstam variables.
We find that the cross-section of \(e^+e^- \to \omega \gamma\) is three orders of magnitude smaller than that of \(e^+e^- \to \gamma \gamma\) from pure QED backgrounds. 
Moreover, BESIII~\cite{BESIII:2022oww} has explored the mono-photon signature in the invisible dark photon search \(\gamma A' \to \gamma + \text{inv}\), where the dominant background indeed comes from the \(\gamma\gamma (\gamma)\) QED states with photon mistagging. In their study, there are fewer than five events in each \(E_\gamma\) bin at high energy. Since the suppression from the \(e^+ e^- \to \omega \gamma\) process is three orders of magnitude smaller in cross-section compared to the QED process \(e^+ e^- \to \gamma \gamma\), we conclude that the photon mistagged background from \(\omega \gamma\) can be safely neglected.

Regarding the irreducible background \(e^+e^- \to \omega Z^*(\bar{\nu}\nu)\), a simple estimation suggests that the background is approximately \(\sim \mathcal{O}(10^{-2})\) of the signal and can thus be safely disregarded. Therefore, we assume zero background events for the \(\omega + \text{invisible}\) analysis. 

Finally, in the analysis of the \(e^+ e^- \to \omega + \text{invisible}\) process, we use the hadronic reconstruction efficiency of the \(\omega\) meson to be \(65\%\) at both BESIII and STCF \cite{BESIII:2018yvu}. 
Together with the cross-section of \(e^+ e^- \to \omega + a\), we obtain the signal events at BESIII and STCF and find that the
projected limits are superseded by the existing NA62 and
Beam Dump limits, so we do not present the estimated invisible channel sensitivity in Fig.~\ref{fig:results}.

\subsubsection{$\omega + {\rm displaced ~vertex}$}

For the axion with an intermediate lifetime, it can decay away from the primary vertex but still inside the detector. Therefore, we consider the signal process \(\omega\,a \to \omega + (\gamma \gamma)\) with a displaced vertex (DV) of diphoton. The BESIII experiment did not conduct such searches, probably due to the spatial resolution issue in reconstructing the diphoton vertex. Fortunately, the STCF offers an excellent position resolution for the diphoton vertex ranging from \(3-6\) mm, and approximately \(1.5-5\) cm along the z-axis \cite{Achasov:2023gey}.
Based on the geometry and position resolution of the STCF, we impose a requirement on the displacement of the diphoton vertex:
\begin{align}
5 \, \text{cm} < d_a < 2.4\,\text{m} ~,
\label{eq:DVcuts}
\end{align}
where \(d_a\) represents the axion decay length in the laboratory frame. Given the rapid decay of \(\pi^0\) with \(c\tau_{\pi_0} \approx 2.6 \times 10^{-6}\) cm, and the distinctive characteristics of displaced diphotons plus an $\omega$ meson, the SM background events can be safely neglected. 
With some basic cuts, the geometric cuts from Eq.~\eqref{eq:DVcuts}, and the hadronic efficiency of \(\omega\), we estimate the signal events at STCF for \(e^+ e^- \to \omega a \to \omega + \text{DV}(\gamma \gamma)\). The limits on the gluon coupling \(c_{gg}\) are shown in Fig.~\ref{fig:results}, labelled as ``STCF (displaced vertex)".

\subsubsection{$\omega + a (\gamma \gamma)$ }

For the axion with a short lifetime, we consider its prompt decay into \(\gamma \gamma\) due to its dominance in the branching ratios.
In practice, we consider \(e^+e^-\to \omega a (\gamma\gamma)\) produced at the \(J/\psi\) resonant pole for BESIII and the future successor STCF, which will have a huge integrated luminosity at the resonant point. To satisfy the prompt decay condition, we conservatively require the axion decay length \(d_a<1\)~cm.

At the \(J/\psi\) pole, besides the ordinary \(\omega + a\) production via an off-shell photon \(\gamma^*\), there is a significant contribution from the \(J/\psi\) meson decay \(J/\psi \to \omega + a\). The resonance \(e^+e^-\to J/\psi\) production runs at \(\sqrt{s}=3.097\)~GeV with a resonant cross-section \(\sigma(e^+e^-\to J/\psi)\approx 3.234\times 10^6\)~pb~\cite{BESIII:2021cxx}. 

Next, we need to determine the decay branching ratio of \(J/\psi \to \omega + a\). 
The \(J/\psi\) meson decay can be thought of as a pair of $c$-$\bar{c}$-quark annihilating into an s-channel off-shell photon, which further transitions into the final states \(\omega a\) through the $a$-$\omega$-$\gamma$ vertex or \(e^+ e^-\) through the QED vertex. 
Therefore, its decay branching ratios have the following relation~\cite{Merlo:2019anv}:
\begin{equation}
    \frac{{\rm BR}(J/\psi\to\omega a)}{{\rm BR}(J/\psi\to ee)} = \frac{m_{J/\psi}^2}{32\pi\alpha}\left\vert c_{\omega\gamma}^{\rm eff}(q^2=m_{J/\psi}^2)\right\vert^2\left[\left(1-\frac{(m_a+m_\omega)^2}{m_{J/\psi}^2}\right)\left(1-\frac{(m_a-m_\omega)^2}{m_{J/\psi}^2}\right)\right]^{\frac{3}{2}} ~,
\end{equation}
where the coefficient \(c_{\omega\gamma}^{\rm eff}(q^2=m_{J/\psi}^2)\) includes the off-shell photon form factor for the $a$-$\omega$-$\gamma$ vertex. The electronic decay branching ratio is \({\rm BR}(J/\psi\to ee)=0.05971\)~\cite{ParticleDataGroup:2022pth}, and we can calculate the signal branching ratio \({\rm BR}(J/\psi\to\omega a) \) accordingly.

For the background estimate of the \(\omega+(2\gamma)\) channel, we refer to a similar axion study through \(J/\psi\) exotic decay at BESIII~\cite{BESIII:2024hdv}. The BESIII study searched for the axion particle through the \(J/\psi\) exotic decay \(J/\psi \to \gamma a \to \gamma (\gamma \gamma)\) with \(a \to \gamma \gamma\) prompt decay. Therefore, our process is very similar to theirs, with the final state \(\omega\) replacing the photon. In their study, they found that the dominant background is \(3\gamma\) from QED processes, rather than photons from hadronic decays ({\it e.g.}, \(\pi_0\) decays), provided the invariant mass of the diphoton (\(\gamma \gamma\)) does not coincide with the resonant mass of \(\pi_0\), \(\eta\), etc.

Therefore, using the VMD model, we expect the background for the \(\omega+a (2\gamma)\) channel to be dominated by the QED-like process \(\omega + 2\gamma\), where one changes one \(\gamma\) to \(\omega\) through the \(\gamma\)-\(\omega\) mixing. We perform the Monte Carlo simulation to estimate this background and calculate the event number in the \(m(\gamma\gamma)\) invariant diphoton mass spectrum. We assume an invariant diphoton mass resolution of 5 MeV and take it as the bin size for the spectrum~\cite{BESIII:2024hdv}. We avoid analyzing the \(\pi^0\), \(\eta\), and \(\eta^\prime\) resonance regions defined within \(m_{\gamma\gamma}\in[0.11,0.16]\)~GeV, \([0.53,0.57]\)~GeV, and \([0.93,0.98]\)~GeV, respectively. In the remaining regions of the spectrum, we are then able to obtain the background number \(N_{\rm bkg}^{\omega+2\gamma}\).

For the signal simulation, the hadronic reconstruction efficiency of the standalone \(\omega\) is taken to be 65\% ~\cite{BESIII:2018yvu}. Furthermore, we perform a MC simulation using \texttt{MadGraph5\_aMC@NLO}~\cite{Alwall:2014hca} to estimate the reconstruction efficiency of the photon pair, applying the following selection cuts on the photons, aligned with those used in Refs.~\cite{BESIII:2022zxr,BESIII:2022rzz}:
\begin{equation}
    E_\gamma > \begin{cases}
        25~{\rm MeV} ,~ \vert\cos\theta\vert < 0.86  \\
        50~{\rm MeV} ,~ 0.86 < \vert\cos\theta\vert < 0.92
    \end{cases} , \quad \Delta R_{\gamma\gamma} > 0.1 ~.
\end{equation}
After that, we obtain the signal event number
\begin{equation}
    N_{\rm sig}^{\omega+(2\gamma)} = \mathcal{L}_{\rm int}\times\sigma(e^+e^-\to J/\psi) \times {\rm BR}(J/\psi\to\omega a) \times {\rm BR}(a\to\gamma\gamma)\times\epsilon_{\omega} \epsilon_{2\gamma} ~,
\end{equation}
where \(\epsilon_{\omega}\) and \(\epsilon_{2\gamma}\) are the detection efficiencies for the hadronic \(\omega\) and the diphoton, respectively.
Comparing the background and the signal event numbers, we set the limits on the gluon coupling \(c_{gg}\) in Fig.~\ref{fig:results}, labelled as ``$e^+ e^- \to \omega \,a (\gamma \gamma)$".

\end{document}